\documentclass[a4paper,11pt]{article}
\pdfoutput=1
\usepackage{NASApub}

\title{A New Interpretation of Einstein's Cosmological Constant}

\author{Thomas L. Wilson}

\affiliation{NASA Johnson Space Center,\\
2101 NASA Parkway, Houston, Texas 77058, U.S.A.}

\emailAdd{thomas.l.wilson@nasa.gov}

\abstract{A new approach to the cosmological constant problem is proposed by modifying Einstein's theory of general relativity, using 
instead a scalar-tensor theory of gravitation.  This theory of gravity crucially incorporates the concept of quantum symmetry breaking.  
The role of the cosmological constant $\lambda$ as a graviton mass in the weak-field limit is necessarily utilized.  Because $\lambda$ 
takes on two values as a broken symmetry, so does the graviton mass -- one of which cannot be zero.  Gravity now exhibits both 
long- and short-range forces, by introducing hadron bags into strong interaction physics using a nonlinear, self-interacting scalar 
$\sigma$-field coupled to the gravitational Lagrangian.}

\begin{document}

\maketitle\flushbottom

\section{Introduction}
\noindent
The question of the existence and magnitude of Einstein$\text{'}$s cosmological constant 
$\lambda$ [1,2] remains one of fundamental significance in our understanding of 
the physical Universe.  The original motivation for introducing $\lambda$ in 
General Relativity [1] 
\begin{equation}
R_{\mu \nu} - \frac{1}{2} g_{\mu \nu} R + \lambda g_{\mu \nu} = -–\kappa T_{\mu \nu}
\end{equation}
\noindent
addressed cosmology and lost much of its appeal after the discovery of an expanding 
Universe.\footnote{In (1.1), $R$ is the spacetime curvature, $R_{\mu \nu}$ is the Ricci 
tensor, $g_{\mu \nu}$ is the spacetime metric, $T_{\mu \nu}$ is the energy-momentum tensor, 
and $\kappa=8\pi G/c^4$ with $\hat{\kappa}=\kappa c^2$ where $G$ is Newton$\text{'}$s 
gravitation constant and $c$ is the speed of light. Metric signature is ($-$,+,+,+).  
Natural units are adopted ($\hbar=c=1$) and $x=(\bf{x},$t).}  However, Einstein$\text{'}$s 
second attempt [2] to reinterpret the meaning of the $\lambda$ term as related to the 
structure and stability of matter has received little attention [3,4,5].  There he used 
$\lambda$ to define a traceless symmetric energy-momentum tensor $T_{\mu \nu}$ which freed 
the field equations (1.1) of scalars, arguing that this contributed to the equilibrium 
stability of the electron.\footnote{There are two source contributions in $T_{\mu \nu}$ 
which can couple to a scalar Spin-0 field:  The trace 
$T=T_\mu^{~\mu}$ and $T_{\mu \nu}^{~~,\mu \nu}$.  Energy-momentum conservation 
$T_{\mu \nu}^{~~;\nu} = 0$ guarantees $T_{\mu \nu}^{~~,\mu \nu} = 0$.  Commas represent 
ordinary derivatives $\partial_\mu$ and semi-colons covariant derivatives $\nabla _\mu$.  
These will be used interchangeably throughout.}  Weyl subsequently justified the 
$\lambda g_{\mu \nu}$ term in (1.1) further by proving that 
$g_{\mu \nu}$, $g_{\mu \nu}R$, and $R_{\mu \nu}$ are the only tensors of second order that 
contain derivatives of $g_{\mu \nu}$ only to second order and only linearly [6].\\
\indent
With the later advent of quantum field theory (QFT), it was recognized that $\lambda$ 
is actually a vacuum energy density [7,8]. There now exists an empirical disparity 
between the universal vacuum energy density in cosmology 
$ ({{\sim 2 \cdot 10^{-3}}~eV})^{4}$ and that in hadron physics, e.g., the bag 
constant $B$ $\thicksim$ ${(146~MeV)}^{4}$, which differ by 44 decimal places.  This 
is known as the cosmological constant problem (CCP), and has come to be described as 
one of the outstanding problems of modern physics [9].  Regardless of its outcome, 
Einstein$\text{'}$s discovery of the vacuum energy density $\lambda$ may possibly be his 
greatest contribution to physics.  Weyl$\text{'}$s observation, however, demonstrates that 
$\lambda$ cannot be carelessly neglected and that the CCP represents a serious 
difficulty with Einstein$\text{'}$s theory.

\subsection{Why Modify Einstein Gravity?}	

The purpose of the present report is to introduce another strategy for addressing this long-standing 
circumstance by modifying Einstein gravity to include an additional scalar field 
$\phi$ that is nonminimally coupled to the Einstein-Hilbert Lagrangian, as in 
scalar-tensor theory.  The cosmological term $\lambda$ in (1.1) will be treated as a 
potential term $\lambda(\phi)$ that is driven by this additional self-interacting 
scalar field $\phi$.\\
\indent
Then borrowing from QFT, the self-interacting potentials 
$\lambda$($\phi$) $\rightarrow$ $U$($\phi$) that have been studied in spontaneous 
[10-12] and dynamical symmetry breaking [13] are obvious candidates for merging 
Einstein gravity (1.1) with $\phi$.  This makes quantum particle physics manifestly 
present in order to address both classical and quantum aspects of the CCP.\\
\indent
Because experiment (to be discussed in \S 4) has clearly 
shown that Einstein gravity is the correct theory for long-distance gravitational 
interactions, that fact will prevail here too.  There is no current experiment 
that can distinguish between Einstein gravity and the modification proposed here in 
this report.  The effect of the JFBD mechanism will only change gravitation at very 
small sub-mm distances such as the GeV and TeV scale of hadron physics, and beyond 
the Hubble radius in cosmology.\\
\indent
We will defer some of the controversial points about inconsistencies in QFT, 
problems with renormalization versus unitarity, and Einstein versus quantum gravity 
to Appendices or later in the text.  In \S 1.2 preliminaries are discussed, while in 
\S 1.3 merging hadrons with gravity is presented.  In \S 2 the scalar-tensor mechanism 
is developed.  In \S 3 the subject of $\lambda$ and graviton mass is addressed, and 
\S 4 will discuss experimental aspects.  Then comments and conclusions 
follow in \S 5.  All assumptions are summarized in \S 5.3.

\subsection{Preliminaries}\

\subsubsection{Symmetry Breaking Potentials $U$($\phi$)}

Examples of symmetry breaking potentials $U$($\phi$) include the quartic Higgs potential 
for the Higgs complex doublet $\phi$ $\rightarrow$ $\Phi$
\begin{equation}
U(\Phi) = - \mu^2({\Phi}^{\dag}{\Phi}) + \zeta({\Phi}^{\dag}{\Phi})^2  \quad ,
\end{equation}
\noindent
where $\mu^2>0$ and $\zeta>0$.  (1.2) has minimum potential energy for 
$\phi_{min} = \frac{1}{\sqrt{2}} {0 \choose \nu}$ with $\nu = \sqrt{\mu^2/ \zeta}$.  Viewed as a 
quantum field, $\Phi$ has the vacuum expectation value $<\Phi>=\Phi_{min}$.  
Following spontaneous symmetry breaking (SSB), one finds 
$\Phi_{min} = \frac{1}{ \sqrt{2} }{0 \choose {\nu + \eta (x)}}$, 
indicating the appearance of the Higgs particle $\eta$.  In order to obtain the 
mass of $\eta$ one expands (1.2) about the minimum $\Phi_{min}$ and obtains\\
\begin{equation}
U(\eta) = U_o + \mu^2\eta^2 + \zeta\nu\eta^3 + {\frac{1}{4}}\zeta\eta^4   \quad ,
\end{equation}
\noindent
where $U_o = - \frac{1}{4} \mu^2\nu^2$ and $\eta$ acquires the mass 
$m_{\eta} = \sqrt{2 \mu^2}$.

\indent
Another example of such potentials is the more general self-interacting quartic case\\
\begin{equation}
U(\phi) = U_o + \kappa\phi + \frac{1}{2}m^2 \phi^2 + \zeta\nu\phi^3  
+ {\frac{c}{4!}}\phi^4 \quad ,
\end{equation}
\noindent
investigated by [14] to examine the ground states of nonminimally coupled, 
fundamental quantized scalar fields $\phi$ in curved spacetime.  $U_o$ is arbitrary.  
(1.4) is based upon the earlier work of T.D. Lee et al. [15-16] and Wilets [17] 
for modelling the quantum behavior of hadrons in bag theory
\begin{equation}
U(\sigma) = U_o + \frac{d}{2} T^*\sigma + \frac{a}{2}\sigma^2 + {\frac{b}{3!}}\sigma^3  
+ {\frac{c}{4!}}\sigma^4 \quad ,
\end{equation}
\noindent
where $\phi$ $\rightarrow$ $\sigma$ represents the scalar $\sigma$-field as a 
nontopological soliton (NTS).  $U_o=B$ is the bag constant and is positive.  The 
work of Lee and Wilets is reviewed in [18-21].\\
\indent
In all cases (1.2)-(1.5), $U_o$ represents a cosmological term, and all are unrelated 
except that they represent the vacuum energy density of the associated scalar field.  
The terms in $U(\phi)$ have a mass-dimension of four as required for renormalizability.  
In the case of (1.2)-(1.3), it is the addition of the Higgs scalar $\eta$ that makes 
the standard electroweak theory a renormalizable gauge theory.  Also, the electroweak 
bosons obtain a mass as a result of their interaction with the Higgs field $\eta$ 
if it is present in the vacuum.\\
\indent
Note finally that (1.3)-(1.5) all have the same basic quartic form.\\
\indent
In what follows, we will examine (1.2)-(1.5) and relate them to the CCP using (1.5).  
This will be done in the fashion of a modified Jordan-Fierz-Brans-Dicke (JFBD) 
scalar nonminimally coupled to the tensor field $g_{\mu \nu}$ in (1.1). 

\subsubsection{The Classical and Quantum Vacuum Issue}

In order to address the vacuum energy densities associated with the CCP, one speaks 
of vacuum expectation values (VEVs) which in turn require an understanding of the 
vacuum.  The vacuum is calculated in \S 1.2.1 to be the ground state or state of 
least energy of the field $\phi$, while treating it as a classical field.  From its 
globally gauge invariant Lagrangian 
$\mathfrak{L}_{\phi} = \partial_{\mu} \phi \partial^{\mu} \phi –- U(\phi)$, this means 
finding the minima of the potential energy $U(\phi)$ as well as the vanishing of kinetic 
energy terms $\partial_{\mu} \phi \partial^{\mu} \phi = 0.$\\
\indent
On the other hand, in QFT the vacuum is defined as the ground state of all quantum 
fields [e.g., 8].  This is reasonably defined provided one does not introduce a 
gravitational field.  Currently a consistent theory of quantum gravity does not 
exist [22,23], although there have been attempts to examine the VEVs of quantized 
scalar fields on curved backgrounds [e.g., 13,24].  Hence, we will define the QFT 
vacuum as the ground state of all quantum fields that exist in and can interact 
with one another in a gravitational vacuum.\\
\indent
However, the absolute values of VEVs are known not to be measurable or observable 
quantities.  Some can be infinite.  Only the energy differences between excited 
states in QFT are experimentally determinable.  This is true regardless of their 
renormalization and regularization [8].\\
\indent
With respect to curved backgrounds, an additional point of view regarding QFT and gravity 
will be presented here.  In gravitational perturbation theory, the metric field 
$g_{\mu \nu}$ in (1.1) is defined as $g_{\mu \nu}$ = $\eta_{\mu \nu}$ + $h_{\mu \nu}$ 
where $\eta_{\mu \nu}$ is the classical background and $h_{\mu \nu}$ is the perturbation 
(illustrated in Appendix A) or quantum fluctuation.  From this point of view, 
one can use $\eta_{\mu \nu}$ to define the ground-state or zero-point energy of the 
classical gravitational vacuum, noting that the total energy of the Universe represented by 
$\eta_{\mu \nu}$ is constant -– and arguably is zero [25].\footnote{Given that quantum 
fluctuations must arise in classical-plus-quantum gravity at finite temperature, these 
cannot violate conservation of total global energy.  The argument in [25] that the Universe 
represented by $\eta_{\mu \nu}$ has zero total energy means then that the quantum 
fluctuations about $\eta_{\mu \nu}$  in renormalization and regularization field theory 
must average out to zero with respect to the classical $\eta_{\mu \nu}$.}  In the case of 
Friedmann-Lemaitre accelerating cosmology, for example, $\eta_{\mu \nu}$ is a de Sitter 
space with cosmological constant $\lambda_{F-L}\!\sim\!{10}^{-56} ~cm^{-2}$.  The F-L metric as 
$\eta_{\mu \nu}$ will be assumed here, noting that the key word is \textit{assumed}.\\
\indent
Nevertheless, the entire subject of gravitational ground-state vacua is probably 
the least understood of all physics in reaching an ultimate understanding of the CCP and its 
solution (T. Wilson, to be published).

\subsubsection{Why NTS Bags?}
\noindent
There are several reasons for making the soliton bag (1.5) the choice for the scalar 
field.  The principal reason is that it represents something known to exist, the 
hadron, and whose vacuum energy density $B$ has been modelled and studied for the 
past 40 years but never unified with gravity.  One would hope that (1.2) from which 
emerged the Higgs $\eta$ (yet undiscovered) might be used instead of (1.4)-(1.5), 
perhaps by equating $U_o$ in (2a) with $B$.  However, this seems impossible because 
$U_o = -¼ \mu^2\nu^2$ for Higgs is negative definite.  Therefore, the Higgs 
mechanism \textit{per se} cannot solve the CCP.  It has the wrong sign.\\
\indent
Another reason for (1.5) is quark confinement, to rectify the fact that quantum 
chromodynamics (QCD) has no scalar field [18].  The introduction of such a 
self-interacting scalar $\sigma$-field seems natural as a preliminary model for 
confinement in hadron physics.\\
\indent
Finally and of relevance here, $B$ is not a ``bare'' number but rather an effective 
vacuum energy density determined by modelling excited states of all hadrons. \\
\indent
The NTS bag model has been introduced by Friedberg \& Lee (FL) [15,16] 
as an attempt to address the dynamics of the confinement mechanism that embeds quarks 
in the QCD vacuum.  It has the important feature that confinement is the result of 
a quantal scalar $\sigma$-field subject to SSB, as discussed in \S 1.2.1.  Earlier
bag models insert confinement by hand, such as the original static 
boundary-condition models of  MIT [26] and SLAC [27], which are purely 
phenomenological in nature.  There are, nevertheless, problems with the FL model.  
It directly couples the $\sigma$-field to the quarks, breaking chiral symmetry [21].  
And it is a quasi-classical approximation [28].\\
\indent
Wilets et al. [17,19] have addressed these problems with the FL NTS model and have 
extended it to permit quantum dynamical calculations.  Known as their 
chromodielectric model (CDM, hereafter FLW model), this includes 
quark-$\sigma$-field coupling and is chirally symmetric [21].\\
\indent
The newest development in bag theory is the derivation directly from QCD by Lunev \& 
Pavlovsky [29,30] which proposes quark confinement based upon singular solutions of 
the classical Yang-Mills gluon equations on the surface of the bag.  Although this 
solution has infinite energy, more recent higher-order modifications to the pure 
Yang-Mills Lagrangian have produced finite-energy, physical solutions for gluon 
clusters and condensates [30].  Similar changes lead to color deconfinement in 
accordance with the asymptotic freedom of quarks [31].  These developments represent 
decided improvements and are not phenomenological results.  For those that view the 
NTS $\sigma$-field as a phenomenological field, they can pursue (1.4) instead of (1.5) 
provided $U_o$ remains positive.\\
\indent
Here the self-interacting $\sigma$-field can be viewed as the bag mechanism which 
creates hadrons as "bubbles" of perturbative vacuum immersed in a Bose-Einstein 
gluon condensate that conceivably makes up the nonperturbative vacuum in QCD.  It 
arises from the nonlinear interaction of the Yang-Mills color fields with the 
$\sigma$-field, and confines the quarks by permitting the appearance of color 
within the bag.  Condensates are scalars, and of course, are necessarily composite 
fields.  Scalars are also the basis of JFBD gravitation theory.  In what follows, 
we will represent the hadron bag as the cosmological term of a fundamental 
scalar-tensor gravitational field.

\subsubsection{The QCD Plus Bag Lagrangian}
\noindent
The FLW NTS Lagrangian is directly connected with that of QCD, since (1.4) and (1.5) 
are related to the $\phi^4(x)$ model used in QFT for investigating the origin of 
SSB [11], extending [11] to gravitational backgrounds [32], and studying SSB at 
finite temperature [33].\\
\indent
Noting that QCD is a renormalizable field theory for the strong interactions [34], 
its Lagrangian $\mathfrak{L}_{QCD}$ is
\begin{equation}
\mathfrak{L}_{QCD} = \mathfrak{L}_q + \mathfrak{L}_c   \quad ,
\end{equation}
\noindent
to which are added gauge-fixing, ghost, counter, and chiral breaking terms.  
$\mathfrak{L}_q$ is the Dirac contribution for quarks and $\mathfrak{L}_C$ is the 
color contribution for gluons
\begin{equation}
\mathfrak{L}_q = \bar{\psi}(i\gamma^{\mu}D_\mu - m)\psi   \quad ,
\end{equation}
\begin{equation}
\mathfrak{L}_C = - \frac{1}{4}F_{\mu \nu}^c F_c^{\mu \nu}   \quad ,
\end{equation}
\noindent
with the gauge field tensor
\begin{equation}
F_{\mu \nu}^a = \partial_{\mu}A_{\nu}^a - \partial_{\nu}A_{\mu}^a 
+ g_sf_{bc}^a A_{\mu}^b A_{\nu}^c   \quad ,
\end{equation}
\noindent
where $D_\mu$ represents the gauge-covariant derivative, $m$ the flavor matrix for 
quark masses, and $g_s$ the strong coupling constant.  Use of the covariant 
derivative introduces the quark-color interaction $\mathfrak{L}_{qC}$.  Note again 
that there is no scalar field in QCD (1.6) [18] - but only quark fields $\psi$ and a 
set of eight ${SU}_3$ color gauge fields $F_{\mu \nu}^c$ with structure constants 
$f_{bc}^a$.  The Higgs in (1.2)-(1.3) is a scalar in electroweak theory, not QCD.\\
\indent
The $\sigma$-field has been described as a scalar gluon field [18], and it obviously 
must be coupled to the gluons in $\mathfrak{L}_C$  and quarks in $\mathfrak{L}_q$.  
In the case of $\mathfrak{L}_C$, this is done using a dielectric coupling coefficient 
$\epsilon(\sigma)$.  That then relates $\sigma$ to the gluon condensate in the 
physical vacuum containing virtual excitations of quarks and other objects.  This 
is accomplished by adding to QCD in (1.6) the $\sigma$-field itself 
$\mathfrak{L}_\sigma$
\begin{equation}
\mathfrak{L}_\sigma = \partial_{\mu}\sigma \partial^{\mu}\sigma - U(\sigma)   \quad ,
\end{equation}
\noindent
consisting of a kinetic term and the self-interaction quartic potential (1.5) in the 
form 
\begin{equation}
	 U(\sigma) = B + \frac{a}{2}\sigma^2 + \frac{b}{3!}\sigma^3 
+ \frac{c}{4!}\sigma^4   \quad ,
\end{equation}
\noindent
where $a,b,c$ are coefficients chosen such that each term in (1.11) has a 
mass-dimension of four (in natural units $\hbar=c=1$), knowing that $\sigma$ has 
dimension one set by the kinetic term in (1.10).  $B$ is the energy density that 
accounts for the non-perturbative QCD structure of the vacuum, measuring the energy 
density difference between the perturbative vacuum (inside the bag) and the true 
nonperturbative ground state QCD vacuum condensate  (outside the bag) [35].  Also 
the fermion-scalar interaction term $\mathfrak{L}_{q,\sigma}$ can be added
\begin{equation}
\mathfrak{L}_{q,\sigma} = - f( \bar{\psi} \psi)   \quad ,
\end{equation}
\noindent
which breaks chiral invariance because $f=f(\sigma)$ is an effective mass added to 
(1.7).  The collective NTS Lagrangian, then, is
\begin{equation}
\mathfrak{L}_{NTS} = \mathfrak{L}_q + \epsilon(\sigma)\mathfrak{L}_C + \mathfrak{L}_\sigma + \mathfrak{L}_{q,\sigma}   \quad ,
\end{equation}
\noindent
which is the standard $\mathfrak{L}_{QCD}$ of QCD supplemented by the nonlinear scalar 
$\sigma$-field and a possible chirality breaking interaction [19],
\begin{equation}
\mathfrak{L}_{NTS} = \mathfrak{L}_{QCD} + \mathfrak{L}_\sigma + \mathfrak{L}_{q,\sigma}  \quad .
\end{equation}
\noindent
$\epsilon(\sigma)$ is the color-dielectric function which depends upon the 
$\sigma$-field [17] and whose form assures color confinement.  In the exact gluon 
limit, $\mathfrak{L}_{NTS}$ $\rightarrow$ $\mathfrak{L}_{QCD}$ because one expects 
$f$ $\rightarrow$ $0$ and $\epsilon$ $\rightarrow$ 1 as the $\sigma$-field decouples 
from the problem [36].\\
\indent
In $\mathfrak{L}_{NTS}$, there are only the scalar field $\sigma$ and the quark fields 
$\psi$ which are a color triplet with $F$ flavors along with the colored gauge gluons.  
Not shown in (1.13) and (1.14) are the counter terms.\footnote{The renormalization of loop 
diagrams for the gauge vector and quark fields requires counter terms, such as given in 
Ref. [15, \bf{D16}].}  The FLW model has been briefly summarized [18,19] and bags reviewed 
[37,20,21-Mosel] elsewhere.\\
\indent  
Interpreting the $\sigma$-field as arising from the nonlinear interactions of the 
color fields in a gluon condensate, while the gluons are also represented separately 
in the Lagrangian, may represent double counting.  This is to be avoided, but does 
not influence one and two gluon exchange [17].\\
\indent
One point of the present study is that $B$ in (1.11) obviously must be related to 
$\lambda$ in (1.1), as $\lambda=\lambda(B)$.  This fact is ignored in the FLW NTS 
model.  Hence $\mathfrak{L}_\sigma$ must also be coupled with a satisfactory 
gravitational Lagrangian relating $U(\sigma)$ in (1.5) and (1.11) to (1.1) since 
gravitation is the presumed origin of the classical vacuum energy density [1,2].\\  
One must guard against over-counting, at any level.  The vacuum energy density can 
only be introduced once, not both in (1.1) and (1.11).  We will show how these merge 
into the same thing by adopting a modified energy-momentum tensor on the 
right-hand-side of (1.1).\\
\indent
The important consequence will be that the $\sigma$-field will emerge as the scalar 
component of the gravitational field in scalar-tensor theory.

\subsection{Gravity And Hadrons}	

It appears that the thought of merging or unifying the NTS bag in hadron physics with 
Einstein or quantum gravity has not occurred to anyone.  Minimal coupling [38] is not 
enough because it does not eliminate the inconsistency of double-counting $\lambda$ in both 
(1.1) and (1.11).  Precluding this is another goal of the present report.\\
\indent
Following Einstein$\text{'}$s work on $\lambda$$\text{'}$s place in unification [2], Dirac 
used an elementary bag theory [39] to address the structure of the electron too.  Hence 
the subject is not really a new one.  However, the notion of radiation- or quantum-induced 
symmetry breaking [11-13] did not exist at the time.  Hopefully, the strategy here will 
interrelate QCD, bag theory, and gravitation in a meaningful way.\footnote{That $\lambda$ and 
$g_{\mu \nu}$ may be weak and can be neglected in QFT outside the region of the hadron in 
particle physics, does not necessarily mean that they can be neglected inside the hadron 
–- particularly if they play a role in the symmetry breaking phase transition of \S1.2.1.}            

\subsubsection{Consequences}

\textit{First consequence}.  Such a merger means that $\lambda$ becomes a function of 
the $\sigma$-field, $\lambda=\lambda(<\sigma>)$, whereby $\lambda$ represents a broken 
symmetry and takes on two different vacuum expectation values, one inside and one 
outside the hadron bag:  $\lambda(\sigma) = \lambda_{Bag} = \hat{\kappa} B$ in the 
hadron interior and $\lambda(\sigma) = \lambda_{Ext.} \equiv \Lambda$ in the exterior.  These 
two vacua are equivalent to the perturbative and nonperturbative vacua respectively in 
QCD.\\
\indent
Einstein gravity cannot do this.  (1.1) contains only a single-valued 
$\lambda$, whereas QCD has two vacua.  The strategy here will become evident in the 
presentation of Figure 1 in the section that follows (\S 1.3.2).\\
\indent
In this scalar-tensor model, the $\sigma$-field is both a color condensate scalar in 
bag theory and a gravitational scalar by virtue of its role in JFBD theory.  It 
couples attractively to all hadronic matter in proportion to mass and therefore 
behaves like gravitation as the scalar component of a Spin-0 graviton but with JFBD 
scaling.\\
\indent
It can be coupled to the metric tensor $g_{\mu \nu}$ of gravitation in several ways.  
As usual, the hadron bag constant $B$ in (1.11) is a function of chemical potential 
$\mu$ and temperature $T$, as $B=B(\mu,T)$, when finite temperature and 
symmetry restoration are considered.  The functional relationship between $B$ and 
$\lambda$ as a function of $\mu$ and $T$ will be determined later  (\S 2.5).\\ 
\indent
\textit{Second consequence.}  The graviton must acquire a mass when $\lambda \neq 0$, 
due to the relationship between $\lambda$ and graviton mass $m_g$.  Details are 
presented in Appendix A [40-60] in the weak-field approximation.\\
\indent
Because of the connection between $\lambda$ and graviton mass $m_g$, unitarity can 
be broken in quantum gravity when $m_g \neq 0$ due to too many Spin-2 helicities.  
This is brought about by the appearance of ghosts and tachyons which are related to 
the propagation of too many degrees of freedom (\S 3.3.2 later).  Since a ghost has a 
negative degree of freedom, more ghosts must be introduced due to perturbative 
Feynman rules that over-count the correct degrees of freedom [61].  The purpose of 
quantum gravity is to straighten all of this out.  To date, that has not been 
possible except for special cases.\\
\indent
Fujii \& Maeda [62, \S 2.6], however, have shown that the scalar $\sigma$-–field of 
scalar-tensor theory couples naturally to the Spin-0 component of $g_{\mu \nu}$.  
There is no problem with degrees of freedom here.  See also the summary in Appendix 
A.2, and [52] as well.

\subsubsection{Gravity and QCD Vacua}

Constructed in flat Minkowski spacetime, the QCD vacuum of particle physics is not an 
empty state but rather is a complicated structure with a temperature-dependent finite 
energy.  Adding gravity would appear to make it even more complicated.\\
\indent
The VEV of the bilinear form for quarks $\bar{\psi}\psi$ distinguishes between the two 
vacua involved here
\begin{equation}
_{QCD}\!<0|\bar{\psi}\psi|0>_{QCD}  ~<~0  \qquad ,
\end{equation}
\begin{equation}
_{Pert}\!<0|\bar{\psi}\psi|0>_{Pert} ~=~0  \qquad ,
\end{equation}
\noindent
where the true nonperturbative QCD vacuum (1.15) creates a pressure that prevents the 
appearance of quarks.  Hadrons are represented by the second VEV (1.16), as a bag of 
perturbative vacuum occupied by quarks and gluons.  These bags or bubbles (1.16) appear 
in (1.15) because of a phase transition occurring in (1.5).\footnote{If one requires that 
the condensates (1.15)-(1.16) must have the same dimensionality (four) as the gluon 
condensate and the Lagrangian, then a quark mass also appears in these equations 
(e.g. [37], p. 366).}\\
\indent
One of the principal dynamic characteristics of the QCD ground state is the bag 
constant $B$.  This is found from the energy difference between (1.16) and (1.15), derived 
as $B_{YM}$ from the Yang-Mills condensate [35].  However, this picture is entirely 
based upon flat Minkowski spacetime.\\
\indent
\begin{figure}[ht]
\begin{center}
\leavevmode
\includegraphics[width=0.8\textwidth]{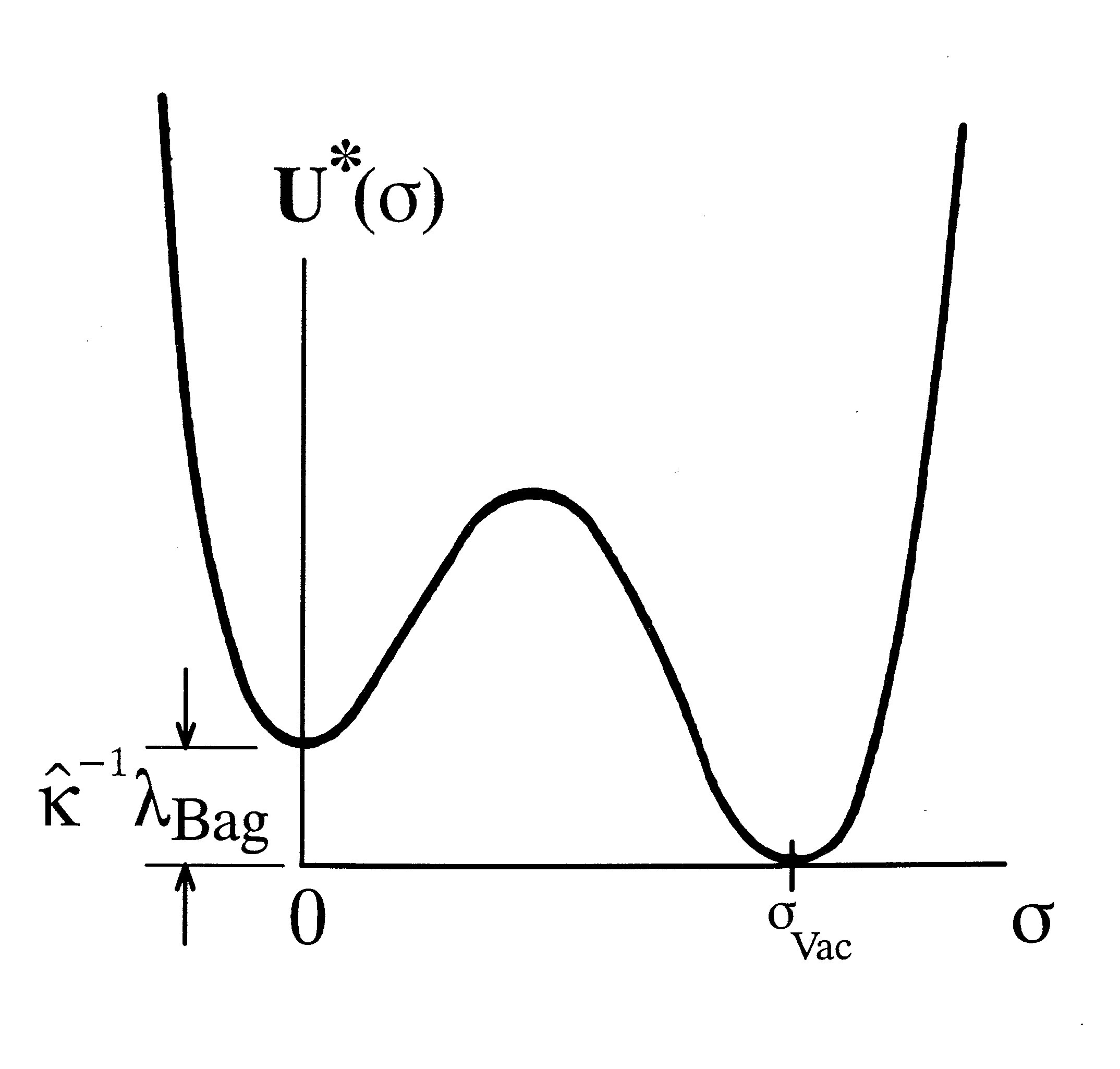}
\end{center}
\caption{In the scalar-tensor model, the cosmological constant $\lambda\!=\!\lambda(\sigma)$ 
has two values because the scalar $\sigma$-field has undergone a phase transition and 
breaks the symmetry of the vacuum, creating two vacuum states.  Inside the hadron, 
$\nu\!=<\!\sigma_{Bag}\!>=\!0$ and $\lambda\!=\!\lambda_{Bag}$, defining $\nu$ as
$\nu\!\equiv<\!\sigma\!>$.  Outside the hadron where $\nu\!=<\!\sigma_{Vac}\!>$, the 
gravitational ground-state energy density of the vacuum $E_{vac}$ is defined by the 
background metric $\eta_{\mu \nu}$ with $\lambda\!=\!\Lambda_{F-L}$ for the 
Friedmann-Lemaitre accelerating Universe.  Both are a de Sitter space.  The hadron bag 
constant $B\!=\!\hat{\kappa}^{-1}\lambda_{Bag}$ is the scale set by $U^*(\nu=0)$.  Eq. (2.6) 
means that $\lambda\!=\!\hat{\kappa}U^*(\nu)$, or $\lambda_{Bag}\!=\!\hat{\kappa}B$ in 
the interior because $\nu\!=\!0$ there.}
\end{figure}
To introduce curvature in the vacuum, one defines gravity as 
$g_{\mu \nu}$ = $\eta_{\mu \nu}$ + $h_{\mu \nu}$.  Necessarily the background 
$\eta_{\mu \nu}$ must be assumed (\S 1.2.2) in order to define the ground-state energy 
for classical gravity where there are no fluctuations ($h_{\mu \nu} = 0$).  Again, for this 
study that will be the accelerating F-L de Sitter space used in cosmology [63].\\
\indent  
Now with reference to Figure 1, define $\nu\!\equiv\,<\!\sigma\!>$.  The ground-state 
energy density occurs at the value $\nu\!=<\!\sigma_{vac}\!>=\!\nu_{vac}$.  This 
corresponds to the VEV $E_{vac}(\nu_{vac})\!=\!U^*(<\!\sigma_{vac}\!>)\!=\!U^*(\nu_{vac})$, 
illustrated by the horizontal axis in Figure 1. It represents the nonperturbative QCD 
vacuum (1.15) external to the hadron in the presence of $\eta_{\mu \nu}$.  This is the 
zero-temperature point where $h_{\mu \nu}$=0.  Similarly at $\nu\!=0$, the VEV 
$E_{vac}(<\!\sigma\!>)=U^*(\nu\!=0)$ represents the perturbative vacuum (1.16) with 
bag constant $B=\hat{\kappa}^{-1}\lambda$, also illustrated in Figure 1.  This figure 
portrays the scalar potential (1.5) used by Creutz [64] and others [65,66] in flat 
Minkowski space.  Varying the parameters such as those of (1.5) [64], one can recover the 
basic bag model of [26] under certain limited circumstances.\\
\indent
Note that $\eta_{\mu \nu}$ is the F-L metric which includes the experimentally 
measured cosmological term $\Lambda_{F-L}$ $\sim$ $10^{-56} cm^{-2}$ corresponding to 
a vacuum energy density $(\sim 2 \cdot 10^{-3}~eV)^4$.  Next note that $E_{vac}$, the 
bag potential function $U^*(\sigma)$, and the value of 
$B = \hat{\kappa}^{-1} \lambda_{Bag}$ in $U^*(\sigma)$ are not observables.  Also, 
$B$ is nonnegative.

\section{The scalar-tensor model \& hadron confinement}

\subsection{Summary of the Problem}

Now we need a brief digression on Einstein gravity (1.1) and how to introduce $g_{\mu \nu}$ into 
hadron physics.  The gravitational field equations (1.1) follow from the classical Einstein-Hilbert 
(E-H) action
\begin{equation}
S = -\frac{1}{2}\kappa^{-1}\!\int d^4 x \sqrt{-g} (R-2\lambda)  \quad .
\end{equation} 
\noindent
where the associated Lagrangian is $\pounds_{EH} = – ½ \kappa^{-1}\sqrt{-g} (R –- 2\lambda)$, 
$g = det ~g_{\mu \nu}$, and the slash in $\pounds$ means that 
$\sqrt{-g} \neq 1$ (i.e., it is not flat Minkowski spacetime).\footnote{This simplifying 
notation is inspired by Feynman$\text{'}$s slash notation used in the Dirac equation.}  
Typically in particle physics, the spacetime is flat, $\sqrt{ -g } \rightarrow 1$, 
and Lagrangians are represented as $\mathfrak{L}$.  Note that a term -\!$ \int d^4 x \lambda$ in 
action (2.1) or -\!$\int d^4 x B$ in (1.11) is not invariant under coordinate 
transformations $x$ $\rightarrow$ $x'(x)$.  The presence of gravity plus Einstein$\text{'}$s 
principle assumption of general covariance or coordinate invariance regarding (1.1) and 
(2.1) requires that such a term be modified to $–\int d^4x \sqrt{(-g)} \lambda $ in 
order that $\pounds$ behaves as a scalar and is a gauge invariant expression.  As a 
result, $\pounds_{EH}$ becomes an infinite series in the metric (graviton) field 
$g_{\mu \nu}$ due to the presence of $\sqrt{-g}$ and $g^{\mu \nu}$ which is the 
inverse of $g_{\mu \nu}$, a property that does not happen in flat spacetime.  In a 
weak-field expansion $g_{\mu \nu}$ = $\eta_{\mu \nu}$ + $h_{\mu \nu}$ about a space 
$\eta_{\mu \nu}$, one has [22]
\begin{equation}
\sqrt{(-g)}=1 + \frac{1}{2}h_\mu^{~\mu} + \frac{1}{8}h_{\mu}^{~\mu} h_\nu^{~\nu} - \frac{1}{4}h_{\mu\nu}h^{\mu\nu} + O(h^3)  \quad .
\end{equation} 
\noindent
This is not a cosmological term because it is dimensionless, but it illustrates what 
begins to happen in quantum gravity for an action (2.1) in curved spacetimes rather 
than flat space.  Obviously, general covariance is broken when 
$\pounds \!\rightarrow \!\mathfrak{L}$.\\
\indent
To restore gravity to the theory of hadrons, several things are required.  The first 
procedure is to introduce minimal coupling by requiring all Lagrangians 
$\mathfrak{L} \!\rightarrow \!\pounds$ while exchanging $g_{\mu \nu}$ for 
flat Minkowski metric terms such as in the electroweak theory 
$\mathfrak{L}_{EW} \!\rightarrow \!\pounds_{EW}$ of the 
SM $\mathfrak{L}_{SM} \!\rightarrow \!\pounds_{SM}$, and hadron theory 
$\mathfrak{L}_{NTS} \!\rightarrow\!\ \pounds_{NTS}$.  Then with 
appropriate gravitationally covariant derivatives, general covariance is restored - 
which guarantees that Christoffel connections appear in the derivatives of (1.9) and 
(1.7) so that Yang-Mills gluons and spinors such as quarks will follow geodesics.  
To this, one adds the postulate of universal coupling [38] which amounts to 
Einstein$\text{'}$s principle of equivalence.  The second, more serious step is the 
introduction of a scalar field, as already discussed in \S 1.2.  The technique is 
developed in Appendix B [67-73].\\
\indent
Several things are already apparent about (2.1).  First, it is notoriously 
nonrenormalizable in the conventional sense of QFT on curved backgrounds and quantum 
gravity because the Newtonian coupling constant G in $\kappa$ has a negative mass dimension
(-2) in four dimensions.  (1.1) and (2.1) are also intrinsically 
nonlinear and are not always subject to perturbative methods, much the same as for 
nonperturbative QCD.\\
\indent
Stelle [74] successfully showed that higher-order terms are in fact renormalizable, 
along with other researchers [75-80] as discussed in Appendix C for further reference 
[74-90].  However, the price one pays is loss of unitarity –- noting that the only 
important example of a theory that is both renormalizable and unitary is 
$\mathfrak{L}_{SM}$ for the SM of particle physics, provided of course that the 
effects of gravity are not included.\\
\indent
In contrast to the higher-derivative approach of Stelle, a method of ghost-free 
gravitational Lagrangians has been found to re-establish unitarity by including terms 
quadratic in curvature as well as torsion [91].  Because of the spin vierbein 
connection in this technique, it may play a role in the unitarity of the Yang-Mills 
portion of QCD that follows here in (14).  Nevertheless, a consistent theory of 
quantum gravity [22,23] or QFT has not yet been formulated (App. C.2) [81-90].\\
\indent
\textit{Summary Issue.}  Another issue about $\pounds_{EH}$ in (2.1) is that 
the vacuum energy density $\lambda$ in (2.1) and $B$ in (1.11) are two very different 
numbers.  The problem here is to connect them in a genuine way.  The first has 
mass-dimension two and the second four.  Hence, there is an inconsistent 
dimensionality of the Lagrangians (1.13) and (2.1) regarding vacuum energy density.  
And again, introducing $B$ in (1.11) and $\lambda$ in (1) constitutes double counting 
which affects their renormalization loop equations.  This will be fixed by 
identifying $\hat{\kappa}^{-1}\lambda$ in (2.1) as the first term of $U(\sigma)$ 
appearing in JFBD theory (Appendix B).  That is, the bag is re-interpreted as a 
potential representing the cosmological term in gravitation theory.

\subsection{Nonminimal coupling of quark bags with gravity}

At the outset, quark confinement in the form of relativistic bags has one 
distinguishing feature as opposed to elementary particle theory.  This feature even 
existed when the electron problem was posed in [2].  The bag is an extended, composite 
object subject to nonlocal dynamics.  This has been pointed out by Creutz [64] while 
examining hadrons as extended objects for the bag model in [26].  Perturbation theory 
is not applicable.  Hence in the presence of confinement, commonly accepted principles 
for point-like particles in QFT such as analyticity of scattering amplitudes are 
called into question.  This must be kept in mind as a possible way around the loss of 
unitarity mentioned in \S 2.1 when strong fields are involved in nonlocal 
strong-interaction hadron physics.  Unitarity may not be required or even possible for 
composite hadron models, although it might be restored using the ghost-free Lagrangian 
methods with torsion just mentioned,\footnote{Torsion is a natural change since it relates 
to the spin connection coefficients that will later appear in (2.10)-(2.13).} or by adopting the 
method in [92].  The Stelle Lagrangian in (B.1) of Appendix B.1 can replace (2.1) in this 
present report in a way that seems acceptable, recognizing that loss of unitarity continues 
to plague quantum gravity at this time.\\
\indent
In order to couple the NTS model in (1.6)-(1.14) with gravity, the total action for 
gravitation, matter, and gravitation-matter interaction is assumed to be
\begin{equation}
S = S_{Gravity}  + S_{matter} + S_{G,m}  \quad .
\end{equation} 
\noindent
Matter will be limited to NTS ($\pounds_{m}$=$\pounds_{NTS}$), 
excluding the electroweak theory $\pounds_{EW}$ of the SM and the Higgs 
fields for this study.  Nonminimal coupling will be used in the sense of 
scalar-tensor theory and the standard additional energy-momentum tensor 
$T_{~\mu \nu}^{\sigma}$ will be added for the $\sigma$-field, with details in 
Appendix B.\\
\indent
Transposing $\lambda$ to the right-hand side\footnote{Geometry in Einstein gravity is 
determined by $g_{\mu \nu}$ - not which side of the equation $\lambda$ is on.} of (1.1) gives 
the scalar-tensor field equations
\begin{equation}
R_{\mu \nu} - \frac{1}{2} g_{\mu \nu} R = -\kappa T^*_{~\mu \nu}  \quad ,
\end{equation} 
\noindent
\begin{equation}
T^*_{\mu \nu} =  T^M_{~~\mu \nu} + T^{\sigma}_{~\mu \nu}  \quad ,
\end{equation} 
\noindent
\begin{equation}
\lambda_{Bag} = \hat{\kappa} B  \quad .
\end{equation} 
\noindent
where now $\lambda=\lambda(\sigma)$ contributes to $T^{\sigma}_{~\mu \nu}$, the matter 
tensor is $T^M_{~~\mu \nu} = T_{\mu \nu}$ in (1.1), and $\kappa T^*_{~\mu \nu}$ is 
conserved by the Bianchi identities.  We will derive (2.5) later [\S 2.3 and (B.35)].  
(2.6) represents the new cosmological bag constant $B=\hat{\kappa}^{-1}\lambda$ 
introduced in this report.  Unlike Einstein [2], we will use the traces 
$T^M = T^{M\mu}_{~\mu}$ and $T^*$ as two of several mechanisms to couple 
gravitation to the NTS $\sigma$-field of quantum bag theory.\footnote{When $\kappa$ is 
variable as $\kappa=\kappa(\sigma)$, then $T_{~~\mu \nu}^{M~~;\nu} = 0$ is assumed by the 
principle of equivalence.  See Appendix B and [69].  The theory can proceed along two 
directions at this point, conserving $T^{\sigma}_{~\mu \nu}$ as well since 
$T^{*~~;\nu}_{~\mu \nu}=0$ in (2.4) and (2.5).  The other option is to conserve only 
$T^*_{~\mu \nu}$ and forgo the principle of equivalence, which will not be addressed here.  
The observation to make is that the vacuum energy density is a component of the potential 
$U(\sigma)$ in (1.11), whereby $\lambda=\lambda(\sigma)$ which means 
$\lambda=\hat{\kappa}U(<\sigma>)$.  This amounts to moving $\lambda$ about within the 
Lagrangian $\pounds=T-U$ for $S$ in (2.3).}\\
\indent  
Notice that the mass-dimension problem discussed in \S 2.1 has now been solved in (2.6).  
Both sides of this equation have mass-dimension two.\\
\indent
We want to determine $T^{\sigma}_{~\mu \nu}$ in (2.5) by introducing the self-interacting 
scalar potential $U(\sigma)$ and relating it to the origin of $\lambda_{Bag}$ in (2.6).  
The interaction term is $\pounds_{G,m}$=$\pounds_{G,\sigma}$.  
The Lagrangian for (2.3) now is
\begin{equation}
\pounds = \pounds_{NTS} + \pounds_{G} + \pounds_{G,\sigma}  \quad ,
\end{equation} 
\noindent
where the original Einstein-Hilbert gravitation term $\pounds_G$ and the 
NTS contribution in (2.7) on a curved background are
\begin{equation}
\pounds_{G} = -\frac{1}{2}\kappa^{-1}R  \quad ,
\end{equation} 
\noindent
\begin{equation}
\pounds_{NTS} = \pounds_{Q} + \pounds_{\sigma} + \pounds_{q,\sigma} + \pounds_{C}  \quad ,
\end{equation} 
\noindent
with Higgs bosons and counter terms neglected.  Eventually 
$\pounds_{\sigma}$ will be removed from (2.9) and made a part of 
$\pounds_{G,\sigma}$ in (2.7).\\
\indent
The critical matter-gravity interaction term 
$\pounds_{G,m} = \pounds_{G,\sigma}$ in (2.3) and (2.7) is viewed 
as the origin of $\lambda$ and conceivably hadron confinement.  It is the symmetry breaking 
term. 

\subsection{The field equations (2.23) \& (2.26)-(2.27)}

In the conventional FLW bag model [19] with covariant derivatives, the quark $\psi$, 
scalar $\sigma$, and colored gluon $C$ terms originally in (1.6)-(1.14) now become 
(2.10)-(2.13) for use in (2.9)
\begin{equation}
\pounds_{q} = \bar{\psi}(i\gamma^{\mu}D_{\mu} - m)\psi  \quad ,
\end{equation} 
\noindent
\begin{equation}
\pounds_{\sigma} = \frac{1}{2}\nabla_{\mu}\sigma\nabla^{\mu}\sigma - U(\sigma)  \quad ,
\end{equation} 
\noindent
\begin{equation}
\pounds_{q,\sigma} = - f\bar{\psi}\sigma\psi  \quad ,
\end{equation} 
\noindent
\begin{equation}
\pounds_{C} = - \frac{1}{4}\mathbf{\epsilon(\sigma)}\mathbf{F_{\mu \nu}F^{\mu \nu}} - 
\frac{1}{2}g_s\bar{\psi}\lambda^cA_c^{\mu}\psi  \quad ,
\end{equation} 
\noindent
where $m$ is the quark flavor mass matrix, $f$ the $\sigma$-quark coupling constant, 
$g_s$ the strong coupling, $\mathbf{F_{\mu \nu}}$ the non-Abelian gauge field tensor, 
$D_\mu$ the gauge-covariant derivative, and $\nabla_\mu$ the gravitation-covariant 
derivative (also in $\mathbf{F_{\mu \nu}}$) with the spin connection derivable upon 
solution of (2.4), defining the geodesics.  $\mathbf{\epsilon(\sigma)}$ is the 
phenomenological dielectric function introduced by Lee et al. [15], where 
$\epsilon(0)\!=\!1$ and $\epsilon(\sigma_{vac})\!=\!0$ in order to guarantee color 
confinement.\footnote{Ultimately a better understanding of strong interaction physics, 
confinement, and gravity may eliminate the need for $\epsilon(\sigma)$.  E.g. [30] does 
this successfully.}  The ${SU}_3$ Gell-Mann matrices and structure factors are 
$\lambda_c$ and $f_{abc}$.\\
\indent
Zel'dovich's original argument [7,8] that the action (2.3) is a vacuum correction for 
quantum fluctuations led Sakharov [93] to expand the gravitational E-H Lagrangian 
$\pounds_{EH}$ in powers of the geometric curvature $R$
\begin{equation}
\pounds(R) = \pounds_o + \pounds_G(R) + \pounds(R^2) + ...  \quad ,
\end{equation} 
\noindent
where $\pounds_o$ is the cosmological term and $\pounds_G$ is (2.8).\\
\indent
The positive contribution of matter and fields $\pounds_m$ in (2.3) is 
thereby viewed as offset by the negative contribution of gravitation and hence 
geometry in (2.14), a sort of back-reaction of the metric.  We interpret 
$\pounds_o$ here as the spontaneous origin of the NTS $\sigma$-field 
whose nonlinear self-interaction breaks the symmetry of the vacuum and creates the 
bag ($B\neq 0$ in (1.11), (2.6), and Figure 1).  The scalar-gravitation field coupling 
can take at least two zero- and first-order forms that relate to $T^*_{~\mu \nu}$ in 
(2.4) and (2.5), 
\begin{equation}
\pounds_{G,\sigma}^{(0)} = -B  \quad ,
\end{equation} 
\noindent
\begin{equation}
\pounds_{G,\sigma}^{(1)} = - \frac{d}{4}T^* \sigma  \quad .
\end{equation} 
\noindent
One can actually picture (2.15) and (2.16) as the first two polynomial terms of 
$\pounds_o$ in (2.14), by defining $U^*(\sigma)$ as –$\pounds_o$
\begin{equation}
\pounds_o = -U^*(\sigma)  \quad ,
\end{equation} 
\noindent
\begin{equation}
U^*(\sigma) = - \pounds_{G,\sigma}^{(0)} - \pounds_{G,\sigma}^{(1)} + U(\sigma)  \quad 
\end{equation} 
\noindent
\begin{equation}
 = B + \frac{d}{2} T^*\sigma + \frac{a}{2}\sigma^2 + {\frac{b}{3!}}\sigma^3  
+ {\frac{c}{4!}}\sigma^4  \quad .
\end{equation} 
\noindent
(2.18)-(2.19) contain the usual $U(\sigma)$ in (2.11), now linked to $g_{\mu \nu}$ in (2.4) 
via $T^*$ and $B$.  The NTS Lagrangian $\pounds_{\sigma}$ in (2.11) can be re-defined to 
include $U^*(\sigma)$ instead of $U(\sigma)$,
\begin{equation}
\pounds^*_{\sigma} = \frac{1}{2}\nabla_{\mu}\sigma\nabla^{\mu}\sigma - U^*(\sigma)  \quad ,
\end{equation} 
\noindent
\begin{equation}
= \pounds_{G,\sigma}  \quad .
\end{equation} 
\noindent

$\pounds_{\sigma}^*$ is $\pounds_o$ plus kinetic terms (momenta) built up from 
derivatives of $\sigma$.  In (2.19),  $a,b,c$ are adjusted to produce two minima 
(Figure 1), one at  $\nu = 0$ and one at a ground-state value $<\sigma_{vac}> = \nu_{vac}$, 
while fitting low-energy hadron properties [64-66,27,19].   The term $T^*\sigma$ is a 
chiral symmetry breaking term used to represent the cloud of pions surrounding the bag 
[66,94,95,19].  $d$ (along with $a,b,c$) adjusts this term by skewing (not tilting) 
the $U^*(\sigma)$ potential and breaking the $\sigma$ $\rightarrow$ $- \sigma$ 
symmetry.\\
\indent
This linear $\sigma$-term ($d\neq 0$) in (2.19) is not necessary to create the bag 
($B\neq 0$), breaks dilatation invariance, and can be dropped ($d=0$).  Furthermore, 
it breaks the renormalizability of $U(\sigma)$ in (1.11) by simple power counting of 
(2.16), $T^*$, and $\sigma$ as discussed in Appendix B,  \S B.5.   If used, $d$ must 
be small enough to preserve the two minima in Figure 1, while slightly skewing the 
broken symmetry about the line $–dT^*\sigma$.  For $d=0$, then $U^* = U$.\\
\indent
Inspired by Sakharov in (2.14), $\pounds_o$ is the origin of $U^*(\sigma)$ 
and the $\sigma$-field.  The term $\pounds_{\sigma}$ has been removed from (2.9) and 
placed in (2.20)-(2.21) and (2.17) as $\pounds_{\sigma}^*$, creating a scalar-tensor 
theory of gravitation via (2.3) and (2.7).  Variation of (2.3), using (2.7)-(2.13) with 
(2.11) replaced by (2.20)-(2.21), gives the field equations (2.4) as well as those for 
$\sigma$ and $\psi$, 
\begin{equation}
\square \sigma = U^{*\prime}(\sigma) + f\bar{\psi}\psi  \quad ,
\end{equation} 
\noindent
\begin{equation}
(i\gamma^\mu D_\mu - m -f\sigma)\psi  \quad ,
\end{equation} 
\noindent
if one neglects the gluonic contribution (2.13).  $\square$ is the curved-space 
Laplace-Beltrami operator, and $U^{*\prime} = dU^*/d{\sigma}$ is  
\begin{equation}
U^{*\prime} = \frac{d}{4} T^* + a\sigma + {\frac{b}{2}}\sigma^2 + {\frac{c}{3!}}\sigma^3  \quad .
\end{equation} 
\noindent
A variant adopts $d=0$ to simplify (2.22) and (2.24) when pion physics is not 
involved.\footnote{That variant of the model couples to the trace $T^M$ instead of $T^*$ in 
(2.16) although it will have the same renormalization problem as $T^*$ when $d \neq 0$.  
See Appendix B.5; also see [70] for an example.}\\
\indent
The $T^{\sigma}_{~\mu \nu}$ contribution in (2.5), which can be "improved" [96], is 
\begin{equation}
T^{\sigma}_{~\mu \nu} = \nabla_{\mu}\sigma\nabla_{\nu}\sigma  -  g_{\mu \nu}\pounds^*_{\sigma}  \quad,
\end{equation} 
\noindent
and is derived in Appendix B.2 as (B.35) and (B.47).  (2.22) and (2.24) are a scalar wave 
equation for $\sigma$ whose Klein-Gordon mass, which is given in (B.38) and follows as 
(2.28), is $m_\sigma=\sqrt{a}$.  Hence $\sigma$ is short-ranged and does not contribute to 
long-range interactions.\\
\indent
$T^M_{~~\mu \nu}$ in (2.5) is the quark and gluon contribution to the matter 
tensor,\footnote{This includes any other form of speculated ``matter.''} and 
$\kappa T^{\sigma}_{~\mu \nu}$ now contains (2.6).  The trace\footnote{Caution must be 
exercised during the variation of a variable coefficient such as $T^*$.  Since $T^*$ is 
known, it must first be substituted into (2.17) before $\delta S=0$ in (2.3), else 
$\delta[\sigma \sqrt{-g}]=0$, and $g=0$ or $\sigma=0$ results.  A similar thing happens 
with $\hat{\kappa}(\sigma)$ in (2.26)-(2.27) and (2.8).} of (2.5) is $T^* = T^M – \sigma_{;a}^2 + 4U^*$, 
with traces for $T^M$, $T^{\sigma}$, and $T^*$ determined in Appendix B.2 and B.5.\\
\indent
The scalar-tensor model permits a number of options $\kappa=\kappa(\sigma)$, one of which 
is developed in Appendix B.2.  There are four pertinent cases: (a) $\kappa(\sigma)$ = 
constant; (b) $\kappa(\sigma) = \sigma$; (c) $\kappa(\sigma) = \sigma^{-1}$; and 
(d) $\kappa(\sigma)$ arbitrary.  (a) is Einstein gravity; (b) turns off 
$T^*_{~\mu \nu}$  in (2.4) [$\kappa(0)=0$] within the bag, leaving an Einstein space 
$R_{\mu \nu}  = – \lambda g_{\mu \nu} $ due to the cancellation 
$\kappa\kappa^{-1}\lambda_{Bag}=\lambda_{Bag}$; (c) is the ansatz originally used by 
Jordan-Fierz-Brans-Dicke [67-69]; and (d) $\kappa(\sigma)$ is any well-behaved 
function of $\sigma$ provided it results in consistent physics, a case that is not 
developed here (although it is discussed in Appendix B.4).\\
\indent
As derived in Appendix B.2 and (B.3) [67-72] following usual treatments, Case (c) 
gives for (2.25) and (2.22)\\
\begin{equation}
(R_{\mu \nu} - \frac{1}{2} g_{\mu \nu} R) = - \frac{8\pi}{\sigma}T^M_{\mu \nu} - 
\frac{\Omega}{\sigma^2}[\sigma_{;\mu}\sigma_{;\nu} - 
\frac{1}{2}g_{\mu \nu}\sigma_{;\alpha}\sigma^{;\alpha}] - 
\frac{1}{\sigma}[\sigma_{;\mu}\sigma_{;\nu} - g_{\mu \nu}\square \sigma] - 
\frac{1}{\sigma}g_{\mu \nu}U^*(\sigma)  \quad ,
\end{equation} 
\noindent
\begin{equation}
\square \sigma = \frac{8\pi}{3+2\Omega}T^* + U^{*\prime} (\sigma) + f\bar{\psi} \psi  \quad .
\end{equation} 
\noindent
where $\Omega = (\kappa_1^{-1} - 3/2)$, and $\kappa_1$ is the source of 
$\sigma$-coupling to the trace $T^M$ traditionally used in JFBD theory.  
$T^{M ~~;\nu}_{~~\mu \nu} = 0$ is assumed in Case (c) (see [70] for an exception).  
$E(\sigma)$ is determined either by the Class A or B auxiliary constraints given in 
Appendix B.3.  The Class A constraint is not renormalizable, while the Class B 
constraint determines $E(\sigma)=1$ by the argument surrounding (B.47).\\
\indent
In brief summary, the scalar field $\sigma$ in (2.27) is now coupled to the trace $T^*$ as opposed to (2.22).  It represents an inverse 
gravitational "constant" $G^{-1}$ or coupling parameter $\kappa^{-1}=(8 \pi G)^{-1}$ 
in (2.26), whose vacuum potential $U^*(\sigma)$ has two ground states that determine 
the vacuum energy density $\lambda(\sigma)$ in (2.6) and Figure 1.  It is in this 
sense that gravitation couples to all physics, because of the $\textit{ansatz}$ in 
(B.3).\\

\subsection{Consequences of the symmetry breaking}

\noindent
The field equations (2.26) and (2.27) are not the traditional JFBD problem in search for a Machian 
influence of distant matter or a time-varying $G$.\footnote{Note that G\"odel has dispelled 
Einstein$\text{'}$s belief that General Relativity is consistent with Mach$\text{'}$s 
Principle [116].  This issue is a common theme throughout Brans-Dicke theory [68].}  (2.27) does 
not have a static solution $G^{-1}=\sigma \sim \Sigma m/r$ [97] because $\sigma$ has only 
short-range interaction by virtue of its mass $m_\sigma=\sqrt{a}$ in (B.38).  To make 
the point, (2.27) can be re-written
\begin{equation}
(\square - m\sigma^2) = \delta U^{*\prime}(\sigma) + \frac{8\pi}{3+2\Omega}T^M + 
f\bar{\psi} \psi  \quad ,
\end{equation} 
\noindent
where $\delta U^{*\prime}$ is the remainder of (2.24) after moving the $a\sigma$ term 
to the left-hand side.  Hence a static solution must have a Yukawa cutoff 
$\sigma \sim (e^{-\mu r})m/r$ where $\mu \sim m_\sigma$.\\
\indent
In addition, note carefully that (2.27) and (2.28) are totally absent from QCD in (1.6) 
–- hence the need for a fundamental scalar boson beyond QCD for hadron theory.\\
\indent
It is true that $G$ is carried along as part of the JFBD method in Appendix B.  However, 
the motivation for a $\sigma$-field coupled to gravitation in (2.27) is to try and solve 
the CCP, not investigate the origin of inertia.\\
\indent
There now exist two characteristic vacuum states for 
$\lambda(\sigma) \rightarrow U^*(\sigma)$ in Figure 1 governed by 
the field equations (2.4), (2.5), (2.19), and (2.26)-(2.27). These are 
\begin{equation}
\lambda(<\sigma_{Bag}>)= \hat{\kappa}B  \quad ,
\end{equation} 
\noindent
inside the hadron bag (1.16), and the ``true'' QCD vacuum $\nu_{vac}=<\sigma_{vac}>$ external 
to the hadron (1.15)
\begin{equation}
\lambda(<\sigma_{vac}>) \equiv \Lambda_{F-L}  \quad ,
\end{equation} 
\noindent
scaled to the gravitational ground state $\eta_{\mu \nu}$  (de Sitter space).\\
\indent
As has been shown in Appendix A.1 for the weak-field limit of $g_{\mu \nu}$, the 
cosmological term $\lambda$ behaves as a graviton mass $m_g$ in (A.21)
\begin{equation}
m_g = \sqrt{2\Lambda_{F-L}/3}  \quad ,  \qquad Hadron ~Exterior
\end{equation} 
\begin{equation}
    = \sqrt{2\lambda_{Bag}/3}  \quad .  \qquad Hadron ~Interior
\end{equation} 
\noindent
In terms of the gravitation constant $G$, using (2.6), the following are 
also true
\begin{equation}
m_g = \sqrt{16 \pi G_N B/3}  \quad ,  \qquad Hadron ~Exterior
\end{equation}
\begin{equation}
    = \sqrt{16 \pi G_{Bag} B/3}  \quad .  \qquad Hadron ~Interior
\end{equation}
\indent
The respective VEDs (2.29)-(2.30) and graviton masses $m_g$ (2.31)-(2.34) are summarized in 
Table I (\S 4).\\
\indent
\begin{figure}[ht]
\begin{center}
\leavevmode
\includegraphics[width=0.8\textwidth]{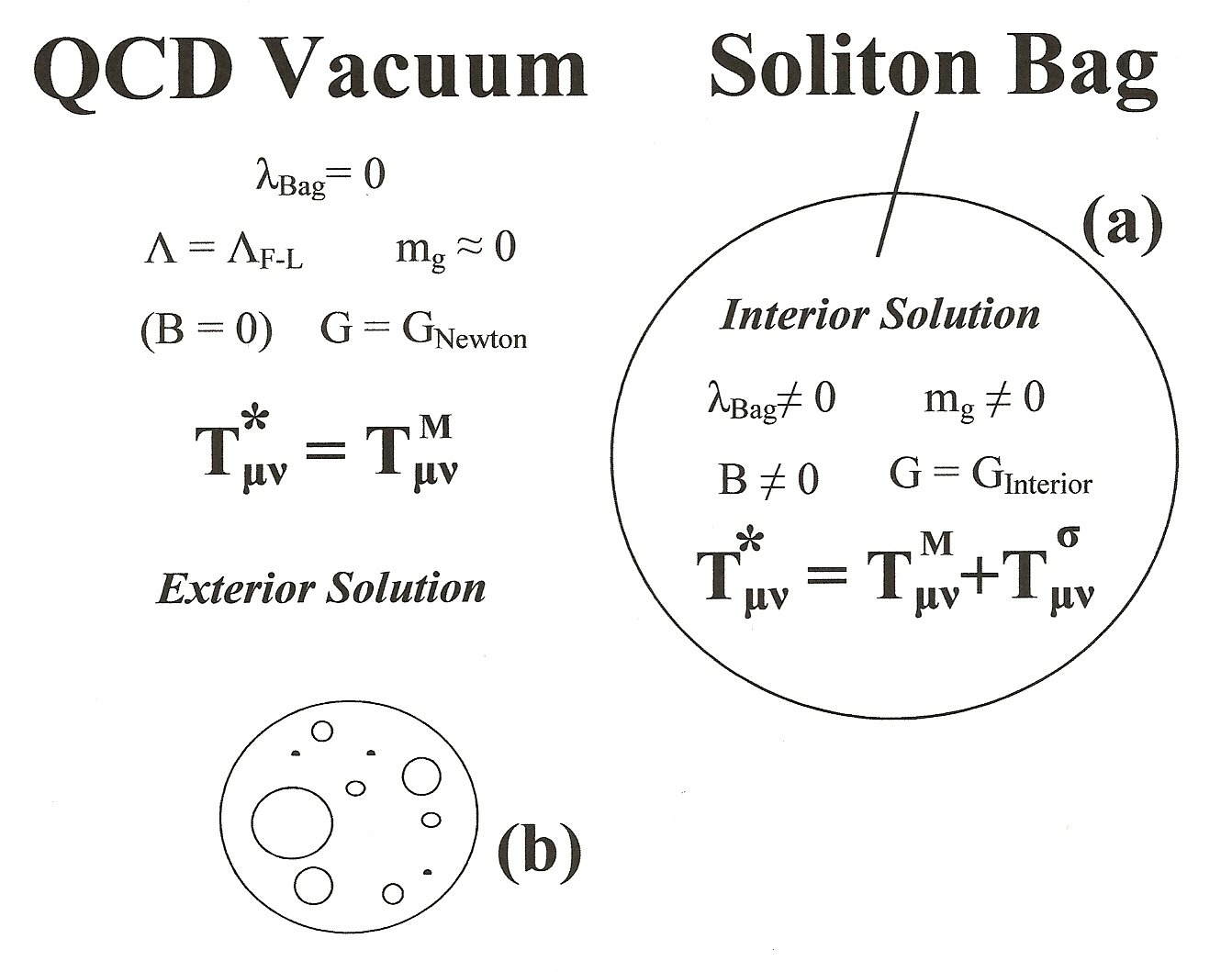}
\end{center}
\caption{The existence of two vacuum states for $\lambda\!=\!\lambda(\sigma)$ 
results in both a short- and long-range gravitational force in hadron physics due to the 
connection between $\lambda$ and graviton mass $m_g$ in the weak-field approximation.  The 
exterior is traditional Einstein gravity for an accelerating F-L Universe, except that the 
graviton has a tiny mass that cuts off at the Hubble radius.  (a) A single soliton bag is 
depicted with a non-zero bag constant. For surface boundary conditions the following 
can be adopted: $F_{\mu \nu}n^\nu\!=\!0$ for gauge fields; and $\psi\!=\!0$ for quark 
fields.  As a problem in bubble dynamics, one uses -$\frac{1}{4}$ $\bf{F}$$^{\mu \nu}$ 
$\bf{F}$$_{\mu \nu}\!=\!–\Sigma J^{-1}\nabla_{\nu}(Jn^\nu)\!+\!B$ for a quark current $J$ 
and surface tension $\Sigma$ [37, Hasenfratz \& Kuti, p. 103].  Alternatively adopt the 
Lunev-Pavlovsky bag with a singular Yang-Mills solution on the bag surface [29,30].  (b) 
The general interior solution is depicted as a many-bag problem using a Swiss-cheese 
(Einstein-Straus) model of space-time with zero pressure on the bag surfaces.}
\end{figure}
\textit{Within the hadron bag.}  Here one has $m_g \neq 0$ due to (2.29) and (2.32).  
Adopting a simplified view of the hadron interior and a bag constant value from one 
of the conventional bag models, the MIT bag [26], $B^{1/4} = 146~MeV$ or $B = 60~MeV 
fm^{-3}$, then $\lambda_{Bag} = \hat{\kappa}B = 2 \cdot 10^{-3}~cm^{-2}$ from 
(2.6).\footnote{Since $1~MeV^4 = 2.3201 \cdot 10^5 g~cm^{-3}$, then $\hat{\kappa}=1.8658 
\cdot 10^{-27} cm~g^{-1}$ or $\hat{\kappa}=4.3288 \cdot 10^{-22} cm^{-2} MeV^{-4}$. Thus 
$\lambda_{Bag}= \hat{\kappa}B=2 \cdot 10^{-13} cm^{-2}$ for $B[g~cm^{-3}] = 2.3201 \cdot 
10^5 B[MeV^4]$.  This assumes $G=G_{Newton}$.}  Using (2.31)-(2.34), a graviton mass $m_g = 3.7 
\cdot 10^{-7} cm^{-1}$ or $7.3 \cdot 10^{-12}~eV$ is found within the bag.  Although this 
appears to represent a Compton wavelength of $m_g^{-1} \sim 3 \cdot 10^6~cm$ or range of 
$\frac{1}{2}m_g^{-1} \sim 1.5 \cdot 10^6~cm$, it is derived from $\lambda_{Bag}$ and is 
only applicable for the interior solution.  This is depicted in Figure 2.  It has no range 
outside of the bag where $\lambda_{Bag}$=0.\\
\indent
A similar calculation for the Yang-Mills condensate $B_{YM} \sim 0.02 ~GeV^4$ [35] 
gives $\lambda_{YM} \sim 8.7 \cdot 10^{-12}~cm^{-2}$ and $m_g \sim 2 \cdot 10^{-6}~cm^{-1}$ 
or $4 \cdot 10^{-11}~eV$, and $\frac{1}{2}m_g^{-1} \sim 2.5 \cdot 10^5 ~cm$.\\
\indent
Regarding $G$, adopting $G_{Bag} = G_{Newton}$ is the conservative assumption to make.  
However, $G_{Bag}$ in (2.34) is a free parameter, independent of $B$.  It has never 
been experimentally measured.  For any $B$ determined in Table I of \S 4, $G_{Bag}$ 
can be anything except zero.  It can re-scale the Planck mass, and therefore 
represents a new way of looking at the hierarchy problem (\S 4.4).  Only experiment 
can determine its outcome.\\
\indent
\textit{External to the hadron}.  This is the nonperturbative QCD vacuum (1.15).  By 
taking the well-known JFBD limit $\Omega \rightarrow \infty$ in (2.26)-(2.27), we in fact 
obtain Einstein gravity (for exceptions see [98]) due to the experimental limits 
given in \S 4.  The small graviton mass $m_g$ in (2.31), on the other hand, results in 
a finite-range gravity whose mass is $m_g \sim 0.8 \cdot 10^{-28}~cm^{-1}$ or 
$1.6 \cdot 10^{-33} ~eV$.  This follows from the vacuum energy density 
${\sim}$ $({{2 \cdot 10^{-3}}~eV})^{4}$ which is equivalent to 
$\lambda \sim 10^{-56}~cm^{-2}$, for the de Sitter background $\eta_{\mu \nu}$ in the F-L 
accelerating Universe [63].\\
\indent
Obviously, $G=G_{Newton}$ in the exterior (Figure 2b).\\
\indent
\textit{Summary.}  The results are as follows.  (2.31) gives a graviton mass 
$m_g \sim 0.8 \cdot 10^{-28}~cm^{-1}$ and a range of $\frac{1}{2}m_g^{-1} \sim 6 \cdot 
10^{27}~cm$ which is approximately the Hubble radius.  That is, gravitation within 
the bag is short-ranged, and gravitation outside of the bag is finite-ranged on the 
order of the Hubble radius.  All of these cases are discussed further in \S 3 as how 
they relate to hadrons, and are summarized in \S 4 as how they relate to 
experiment.\\
\indent
Figure 2 depicts a single soliton bag embedded in the NTS-QCD gluonic vacuum (Fig. 2a), 
as well as for the many-bag, "Swiss-cheese" model of spacetime [99] with 
bags-within-bags which results for multiple hadrons (Fig. 2b).\\
\indent
Clearly the sign of $\lambda$ must be positive (de Sitter space) in (2.31)-(2.34) in order 
that an imaginary mass not be possible.  The latter represents an unstable condition 
with pathological problems such as tachyons and negative probability [41].  (2.31)-(2.34) 
is a physical argument against such a circumstance.

\subsection{Finite temperature effects}

\noindent
The effect of finite temperature $T$ upon $U^*(\sigma)$ is treated in the usual 
fashion [100-102].  The classical, zero temperature potential $U^*(\sigma)$ becomes 
$V^*(\sigma) =  U^*(\sigma) +  V_S(\sigma,T) + V_F(\sigma,T,\mu)$ involving scalar 
$V_S$ and fermionic $V_F$ correction terms for chemical potential $\mu$, by shifting 
$\sigma$ as $\sigma = \sigma^{~\prime} + \nu(T)$.  The result is a temperature-dependent 
cosmological bag parameter [103] $\lambda_{Bag} = \lambda_{Bag}(\mu,T) = 
\hat{\kappa}B(\mu,T)$ which decreases with increasing temperature $T$ until the bag 
in Figure 2 dissolves and symmetry is restored ($B=0$) in Figure 1.\\
\indent 
In such a case and in simplest form [104], the bag model equations of state (EOS) are
\begin{equation}
\epsilon(T) = k_{SB}T^4 + B  \quad ,
\end{equation} 
\begin{equation}
\ p(T) = k_{SB}T^4/3 - B  \quad ,
\end{equation} 
\begin{equation}
k_{SB} = \frac{\pi^2}{30}(d_B + \frac{7}{8}d_F)  \quad ,
\end{equation}
\noindent
where energy density $\epsilon$ and pressure $p$ now have a temperature dependence 
($T\!\neq\!0$).  The Stefan-Boltzmann (SB) constant $k_{SB}$ is a function of the 
degeneracy factors $d_B$ for bosons (gluons) and $d_F$ for fermions (quarks and 
antiquarks).  The absence of the baryonic chemical potential $\mu$ in (2.35) is a 
valid approximation for ongoing experiments involving nucleus-nucleus collisions.  
(2.35)-(2.37) is relevant to quark-hadron phase transitions and the quark-gluon plasma (QGP).

\section{Short-range gravitation and the hadron interior}
\noindent
So far the principal change has been to incorporate the bag constant of hadron physics 
into scalar-tensor gravitation theory, by treating the cosmological term in (1.1) and (2.6) 
as the potential function in (1.5), (2.5), (2.20), and (2.25).  This has resulted in no 
significant experimental change in the hadron exterior.\\
\indent
The hadron interior, however, is a different matter.  As depicted in Figure 1, 
$\lambda \rightarrow \lambda_{Bag}$ has now increased the VED there by 44 orders of 
magnitude.  This is not to suggest that gravity \textit{per se} can compete with strong 
interaction physics in QCD.$^{5}$  What has changed is that gravity is necessarily involved 
in the Lagrangian $\pounds$ of action (2.3) while interacting with the NTS Lagrangian 
$\pounds_{NTS}$ (2.9) which includes QCD.  Details of what gravity does under these 
circumstances have not appeared in the literature, and that is one of the purpose of this 
section (\S 3).  Only if the gravitation constant $G_{interior}$ is experimentally measured 
to be significantly different than $G_{Newton}$ (\S 3.1.2 below) can these calculations 
play an important role in strong interaction physics.\\
\indent
Referring again to Figure 2, the hadron interior ($r\!<\!r_{Bag}$) is inhabited by quarks 
and gluons and is a de Sitter space with a constant $\lambda_{Bag} \neq 0$.  These exist in 
the presence of a scalar field $\sigma(x)$ having mass $m_\sigma=\sqrt{a}$ in (B.38), and 
a graviton field $g_{\mu \nu}(x)$ of mass $m_g = \sqrt{2\lambda_{Bag}/3}$ in (2.32) whose 
helicities will be discussed later (\S 3.3.2).  The physics of the quarks $\psi$ in (2.23) 
and gluon gauge fields $\mathbf{A^\mu}$ in (1.9) participate in the dynamics of bag excited 
states. Neither $\psi$ nor $\mathbf{A^\mu}$ exists in the external solution, with the bag 
surface behaving like that in the Lunev-Pavlovsky gluon-cluster model.\\
\indent  
In the hadron exterior ($r\!>\!r_{Bag}$), nothing has really changed except for the tiny 
graviton mass (2.33) of $\sim\!10^{-33}eV$.  This part of the problem has already been 
solved.  By current experiments (\S 4), it is Einstein gravity about a charged or neutral 
hadron (with a Reissner-Nordstr\"om or Kottler-Schwarzschild solution) for a hadron mass 
$m_h$.  The graviton mass cuts off at the Hubble radius.  For most conceivable 
applications, it is negligible.\\
\indent
The bag \textit{per se} ($r \leq r_{Bag}$) is governed by the short-range tensor 
field $g_{\mu \nu}(x)$ and short-range scalar $\sigma(x)$, their mutual interactions, 
the confined quarks $\psi$ and gluons $\mathbf{A^\mu}$, as well as the energy and 
pressure balance at the bag$\text{'}$s surface ($r\!=\!r_{Bag}$).  Obviously there can be 
surface currents [37] that guarantee quarks and gluons do not exit the bag, as well as a 
bag thickness or skin depth $\delta$ where this takes place [e.g., 64].  Alternatively, 
the Lunev-Pavlovsky gluon-cluster model is equally possible.  There also is a surface 
tension, since the nonperturbative QCD vacuum must offset the negative bag pressure 
created by $B$.  All of these interactions are nonlinear and non-perturbative.\\
\indent  
Finally, the complicated bag surface and thickness $\delta$ with boundary conditions 
need additional comment.  It is important to observe that the zero pressure boundary 
condition (\S 3.1.1 and \S 3.3.2 below) is ``free'' –- an automatic consequence of the 
phase transition in Figure 1 that created the bag.  The bag constant $B$ is either 
$B\!=\!0$ or $B\!\neq\!0$, one of two states.  Only when symmetry restoration is happening 
because $B\!=\!B(\mu,T)$ is actually temperature-dependent (\S 2.5), do more complicated 
dynamics come into play in the boundary condition problem.

\subsection{Classical solutions for the field equations}

\noindent
Here are details addressing solutions for the equations of motion (2.23) and (2.26)-(2.27).  
Classical solutions may be useful in determining the consequences of the present 
investigation should experiment find that $G_{interior}$ has changed significantly as did 
$\lambda_{Bag}$,\\ 
\indent 
Vacuum solutions of the original JFBD equations ($\lambda=0$) have been well 
investigated although all require quantum corrections [105].  The spherically 
symmetric static field for a point mass (with $c=1$)
\begin{equation}
 ds^2 = - e^{2\nu}c^2dt^2 + e^{2\zeta}dr^2 + r^2d\Omega^2  
\end{equation} 
was first examined by Heckmann \textit{et al.} [106] with later studies by Brans 
[107,68], Morganstern [108], Ni [109], Weinberg [69, pp. 244-248], and others [110].  
Exact [107,111] static exterior and approximate [112,113] interior solutions have 
both been discussed, while exact rotating solutions that reduce to the Kerr solution 
when $\sigma \rightarrow 0$ have been found [114].  In addition, conformal 
transformations of solutions in Einstein gravity have been used to generate JFBD 
solutions [72].  Notably, the approximation [112,113] indicates that there are no 
singularities at $r=0$, the center of the sphere.
 
\subsubsection{Boundary conditions in general}

\noindent
A generalized scalar-tensor theory has many similarities with the 
Kottler-Schwarzschild (KS) problem in Appendix A as far as boundary conditions are 
concerned.  The obvious exception is that the scalar field $\sigma$ is regulated by 
a separate equation of motion (2.27)-(2.28) from the Ricci tensor $R_{\mu \nu}$ in (2.26).  
The $\sigma$-field is significant because without it there would be no bag and hence 
no multiple vacua.  Excitations of $\sigma$ in (2.22), (2.23), and (2.24) couple to quark 
$\psi$ excited states in the hadron interior for modelling hadrons.\\
\indent
Spherical symmetry is assumed because it is a very good approximation to many 
physical situations.  With that in mind, the goal is to establish that there exist 
interior solutions of the $\lambda$JFBD scalar-tensor theory assuming a perfect fluid 
whose pressure, mass density, metric, and scalar functions are everywhere finite in 
the bounded region of the hadron and have zero pressure $p=0$ at its surface.
 
\subsubsection{Possible jump conditions}

\noindent
As discussed in \S 2.4, $G$, $B$, and $\lambda$ are now linked in a fundamental way, 
with only one restriction –- relation (2.6).  The two vacuum states in Figure 1 each 
have their own set of these basic parameters.  There are no experimental short-range 
and strong-force measurements within a hadron that guarantee $G$ must equal Newton$\text{'}$s 
constant there.  It is possible that $G \sim \kappa =\sigma^{-1}$ has two states with 
$G_{interior}$ being quite different from $G_{Newton}$ in the hadron exterior as 
depicted in Figure 2.\\
\indent  
This has a direct consequence, related to the standard Einstein limit of JFBD gravity.  
One obtains Einstein gravity as the JFBD coupling constant $\Omega$ goes to infinity 
($\Omega \rightarrow \infty$ with exceptions [98]) and $G \sim \kappa =\sigma^{-1}$ 
becomes constant.  However, $G=constant$ does not mean that $G_{hadron}=G_{Newton}$ 
at the boundary condition interface between the interior and exterior solutions.\\
\indent
Hence, one must be cautious about $G$ in the interior and exterior solutions that 
follow and must await experimental measurements.
 
\subsubsection{Energetics and boundary conditions for the bag}

\noindent
When Einstein introduced $\lambda$ into physics, he created a negative pressure $p$ 
represented by $B \neq 0$ in Figure 1, that is
\begin{equation}
p = - B  \quad .
\end{equation} 
\noindent
The energy-momentum tensor $T_{\mu \nu}$ for a bag is then
\begin{equation}
T_{\mu \nu} = (\rho - B)u_\mu u_\nu - Bg_{\mu \nu}   \quad ,
\end{equation} 
where $\rho$ is the mass density introduced by quarks and gluons, and $u_\mu$ is the 
4-velocity of the assumed isotropic, homogeneous, incompressible perfect fluid in the 
interior.  The latter is not spatially flat.  It is a de Sitter space with 
$R=4\lambda_{Bag}$ containing $N_q$ quarks, at least in the Einstein limit 
$\Omega \rightarrow \infty$ of (2.26) and (2.27).\\
\indent 
For comparison, consider the bag from the point of view of hadron physics.  Also 
treat it as spherical and static.  All quarks are in the ground state, as opposed to 
a compressible bag [115].  Figure 2 represents a simple hadron containing $N_q$ quarks.  
In its simplest form, the hadron mass $M_h$ for bag volume $V= 4\pi r^3/3$ is [115]
\begin{equation}
M_h = kV^{-1/3} + BV  \quad ,
\end{equation} 
\noindent
where on the right-hand-side the first term is the internal quark energy and the 
second is the volume energy.  The volume is determined by pressure balance between 
the internal quarks and the external QCD pressure, as 
\begin{equation}
\frac{\partial M_h}{\partial V} = - \frac{1}{3}kV^{-4/3} + B  \quad .
\end{equation} 
\noindent
$k$ and $B$ are found from experimental values for the proton charge radius and a 
given nucleon mass.  From (3.4) and (3.5), one has for the static bag
\begin{equation}
V = (k/3B)^{3/4}  \quad ,
\end{equation} 
\begin{equation}
M_h = 4(k/3)^{3/4}B^{1/4}  \quad .
\end{equation} 
\indent
Dynamically, boundary conditions are established on the surface of the bag in order 
that the quarks and gluons cannot get out (since free quarks have not been observed).  
Then hadrons are viewed as excitations of quarks and gluons inside the bag. 
Confinement is achieved phenomenologically (inserted by hand) in the MIT bag [26] by 
requiring that there is no quark current flow through the surface of the bag 
(Figure 2, caption).  This results in a nonlinear boundary condition which breaks 
Lorentz invariance.  A similar requirement will be assumed here for examining 
preliminary solutions, rather than address the Lunev-Pavlovsky gluon-cluster model at 
this time.\\
\indent
The boundary condition generates discrete energy eigenvalues $\epsilon_n$ for the 
quarks where
\begin{equation}
\epsilon_n = c_n/r  \quad ,
\end{equation} 
\noindent
and $c_1=2.04$ for the ground state $n=1$.  Assuming that $N_q$ is the number of 
quarks inside the bag, then their kinetic energy is $E_{kin} = N_q \epsilon_n$ while 
the potential energy $E_{pot}$ is $E_{pot}=BV$.  $E_{pot}$ must be subtracted from the 
total bag energy $E$ in order to obtain the total quark energy
\begin{equation}
E - BV = N_qc_1/r  \quad .
\end{equation}
 
\subsubsection{Special case classical solutions}

\indent
The specific results for the classical solutions are presented in Appendix D 
[116-123].

\subsection{The weak-field versus strong-field limit}

\noindent
It has been shown in Appendix A that for the weak-field limit of $g_{\mu \nu}$  in 
\S A.1, the cosmological term $\lambda$ behaves as a graviton mass  
$m_g =\sqrt{2\lambda/3}$ in (A.21).  We have not shown, however, that $\lambda$ 
behaves as a graviton mass in the strong-field case.  See the caution given regarding 
(A.15).  Certainly with higher-derivative and renormalizable Lagrangians such as (B.1) 
or (B.2), it needs to be demonstrated for very strong gravitational fields that the 
$m_g$ term in (A15) still survives as a graviton mass.\\
\indent
Higher derivatives, viewed as momenta, portray high energy and therefore high 
temperatures.  From \S 2.5, symmetry restoration must eventually set in and the bags 
dissolve or disappear.

\subsection{The strong-field case}

\subsubsection{Historical background}
\noindent
The early 1970$\text{'}$s witnessed a sophisticated revival of the old search for unification 
that dates back to Mie [1] and Weyl [3].  For example, Freund introduced a 
Brans-Dicke scalar [120] with unification in mind following an earlier investigation 
of finite-range gravitation [121].\\
\indent  
It also was suggested by Salam $\textit{et al.}$ [124-125] and independently Zumino 
[126] that hadrons interact strongly through the exchange of Spin-2 mesons behaving 
as tensor gauge fields.  The Salam group adopted a bi-metric theory of gravity by 
adding a second set of Einstein equations (1.1) for a new tensor field $f_{\mu \nu}$ 
that would mix with the usual $g_{\mu \nu}$, producing what they called $f\!-\!g$ gravity.  The 
$f_{\mu \nu}$ field described a Spin-2 particle ($f$-meson) bearing a Pauli-Fierz 
mass [56] similar to the method of [126].\\
\indent  
Two cosmological constants were introduced, one for the $f$-field and one for the 
$g$-field.\footnote{One cannot introduce two \textit{ad hoc} constants in classical physics 
that account for the same thing.}  Hadrons were described as two superimposed de Sitter 
"microuniverses" that interacted through $f\!-\!g$ mixing.  These are very similar to 
present-day multiverses or metaverses.  Unlike QFT, Einstein$\text{'}$s theory offers no 
founding principles upon which to define interactions in an $f\!-\!g$ multi-tensor 
system.  So such a scheme is contrived at best.  This is a fact that plagues any 
multi-metric theory.  Metrics must be imbedded using appropriate boundary conditions 
as in Einstein-Straus [99], not superimposed.\\
\indent
$f\!-\!g$ gravity had a lot of problems and fell into disfavor.  Deser [127] established 
many of the difficulties associated with Spin-2 mixing.  Aichelburg [128] showed it 
was impossible to construct a bi-metric tensor gravity theory in the same spacetime 
without losing causality.  The concept of causal metrical structure breaks down due 
to the existence of two propagation cones.  It was also shown that there exist too 
many intractable Spin-2 helicities (seven and eight) [128,129].  And Freund [120] 
showed that the experimentally observed $f$-mesons were not the quanta of a gauge 
field of strong gravity.\\
\indent
As a matter for historical perspective, the gravitation theory presented here was 
found quite independently of multi-metric theories.  It was arrived at while trying 
to solve the CCP.

\subsubsection{Mass and Spin-2 graviton degrees of freedom}
\noindent
Attributing a graviton mass to the region of confinement, the hadron bag, necessarily 
brings up old problems originating at the beginning of quantum gravity (Appendix C).  
The issue is how to reconcile a graviton mass with the interior of a hadron bag.\\
\indent
\textit{Pauli-Fierz Method}.  The traditional method for introducing a graviton mass 
in Spin-2 quantum gravity is that of Pauli-Fierz [56] because it does not introduce 
ghosts and its Spin-0 helicity survives in the massless limit, naturally leading to 
a JFBD scalar-tensor theory of gravitation [52].  Unfortunately, the Pauli-Fierz mass 
$m_{PF}$ is derived from quantum gravity arguments for massive particles having 
integral Spin-2 on a flat background $\eta_{\mu \nu}$.  As with Veltman [41], this 
is simply incorrect [42] if Einstein gravity is the experimentally correct one (\S 4).  
Pauli \& Fierz focus on Lorentz invariance and positivity of energy after quantization.  
However, they totally ignore the cosmological constant ($\lambda=0$), and its 
association with graviton mass in the weak-field limit (Appendix A).  The 
conventional way of working around this oversight is to introduce the Pauli-Fierz mass term 
as a weak-field perturbation $g_{\mu \nu}$ = $\eta_{\mu \nu}$ + $h_{\mu \nu}$ on a curved 
background $\eta_{\mu \nu}$ which is de Sitter space ($\lambda \neq 0$) [58] instead of a 
flat Minkowski space as they assumed for quantum gravity.\\
\indent
\textit{vDVZ Discontinuity}.  Later, the subject of finite-range gravitation resulted 
in the realization of what is known as the vDVZ discontinuity [130,131,132,52].  In 
the linear approximation to Einstein gravity using the Pauli-Fierz mass term 
(App. \S A.2), the zero-mass limit of a massive graviton does not result in the same 
propagator as the zero-mass case.  The consequence is that giving a nonzero mass 
$m_g = m_{PF}$ to a graviton results in a bending angle of light near the edge of the 
Sun that is 3/4 that of Einstein$\text{'}$s value, and the difference may be measurable [52].  
This quantum gravity dilemma is discussed in [131].  Its resolution is making $m_g$ 
small enough and not using perturbative approximations [133].  That is accomplished 
here in the hadron exterior where the free graviton has a tiny mass and a range on the 
order of the Hubble radius.\\
\indent
As for the interior, there is no bending of light experiment that can be performed 
inside a hadron bag (\S 4).  Hence, the vDVZ discontinuity is not relevant to the 
short-range modification of Einstein gravity presented here, because there is no 
massless limit inside a hadron (Figure 2) where $\lambda_{Bag}$ cannot be zero and $m_{PF}$ 
is not introduced.  In fact, the fundamental premise of the scalar-tensor theory is that 
quantum symmetry breaking has resulted in a finite discontinuity in Figure 1 between the 
two vacua.  This results in two discontinuous values of $\lambda$ and one can even 
conjecture that a similar thing happens to $G$ (\S 3.1.2).  A difference in graviton 
propagators inside and outside the bag is to be expected, cautioning that propagators are 
derived from perturbative Feynman techniques that cannot reflect the nonperturbative 
physical properties of confinement and strong interactions.  Again [133], the vDVZ 
discontinuity appears to be an artifact of perturbation theory.\\
\indent  
A key point is that hadron bags are composite objects.  Some physical behavior that 
applies to elementary particle physics may not apply to hadrons.  Recalling the suggestion 
of Creutz [64] that certain fundamental concepts such as unitarity may be called into 
question when discussing composite systems, it may be time to ask similar questions about 
the graviton degrees of freedom inside a hadron bag.  Ultimately the question is how to 
deal with loss of unitarity in a still undefined quantum gravity, and how to admix the 
boson (scalar, gluon, and graviton) degrees of freedom consistently.\\
\indent
\textit{Helicity Properties in the Exterior.}  Recalling the summary for (A.20) of 
Appendix A.2, the result is a well-behaved massive graviton generated by $\lambda$ 
with two transverse helicities obtained in the weak-field approximation.  The Spin-0 
component is suppressed by coupling to a zero-trace $T^M\!=\!0$ energy-momentum tensor 
while the vector Spin-1 components are eliminated by using the gauge $f_\mu\!=\!0$ in 
(A.8).  That method is applicable here in the hadron exterior (2.31).\\
\indent
\textit{Helicity Properties in the Interior.}  For the interior (2.31), the same 
technique can be applied except that one does not suppress the Spin-0 component 
because this couples to the $\sigma$-field constituting the bag in scalar-tensor 
theory (recall \S 1.3.1, [62]).  This appears as coupling to the trace $T^M\!\neq\!0$ in 
(2.27).  Similarly, one argues that the vector degrees of freedom for Spin-1 couple 
to the Yang-Mills gauge gluon fields of QCD –- without need for the gauge $f_\mu=0$ 
in (A.8).\\
\indent
All five degrees of freedom appear necessary for confinement, although as few as four 
have been discussed under other circumstances [57].  The $\sigma$-field and the 
gluons conceptually can interact with $g_{\mu \nu}$ in such a way as not to lose 
unitarity within the bag –- but that appears impossible to prove in the 
nonperturbative environment of confinement with no consistent theory of quantum 
gravity and no experimental data.

\section{Experimental prospects}

\noindent
In this study a $\lambda$-generated graviton mass (2.31)-(2.34) has appeared, with different 
values inside and outside the hadron.  For the case inside the hadron, that will be 
referred to as the confined graviton.  That outside will be called the free graviton.  
The free graviton has a mass $\sim\!10^{-33}eV$ ($\Lambda_{F-L}\!\neq\!0$) with a range 
extending to the Hubble radius since it is scaled to the vacuum energy density 
$({{\sim\!2 \cdot 10^{-3}}eV})^{4}$ or $\lambda_{F-L} \sim\!10^{-56} cm^{-2}$ 
characterizing the F-L accelerating Universe [63, Blome \& Wilson].  In addition, the 
scalar gluon (condensate) $\sigma$-field  has acquired a mass $m_\sigma\!=\!\sqrt{a}$ 
(B.38).  The $\sigma$-field comprises the hadron bag as a composite object, and 
represents the cosmological term as a potential (\S B.1) in scalar-tensor gravitation 
theory, (2.3)-(2.6).\\
\begin{table}[ht]\footnotesize
\caption{Summary of the masses, vacuum energy densities (VED's), and 
$\lambda \text{'}$s in spacetime.  The QCD vacuum (1.15) scales to the VED of accelerating 
F-L cosmology, with a graviton mass whose range is approximately the Hubble radius.}
\centering
\begin{tabular}[ch]{l*{6} {c}r}
\hline
Spacetime 	& $m_g$ & $m_g$ & $m_{\sigma}$ & VED,$B$ & $\lambda$  \\
Region		& $(cm^{-1})$ & $(eV)$ & $(GeV)$ & $(GeV)^4$ & $(cm^{-2})$  \\
\hline\hline
\textit{Hadron Exterior}  \\
~~$\lambda \equiv \Lambda_{F-L}\neq 0$  	& 0.8x$10^{-28}$ & 1.6x$10^{-33}$ & $$ & 2x$10^{-47}$ & 0.7x$10^{-56}$  \\
\textit{Hadron Interior}  \\
~~MIT bag [26]  	& 3.7x$10^{-7}$ & 7.3x$10^{-12}$ & $\sqrt{a}$ & $0.0045$ & 2x$10^{-13}$  \\
~~Y-M cluster [35]		& 2.4x$10^{-6}$ & 4x$10^{-11}$ & $\sqrt{a}$ & $0.02$ & 9x$10^{-12}$  \\
\hline\hline
\end{tabular}
\label{Table}
\end{table}
\indent
These two principal features, a graviton mass and a scalar gluon mass, are summarized 
in Table I.  Both have experimentally observable consequences.  Basic experimental 
findings and limitations are discussed below in \S 4.1 and \S 4.2, while experimental 
consequences for the $\sigma$-field mass $m_\sigma$ are given in \S 4.3.  Those for 
the graviton mass $m_g$ are addressed in \S 4.4 and \S 4.5. And finally, recent 
developments in the dilepton channels of jets at Fermilab are related to a possible scalar 
gluon condensate in \S 4.6.

\subsection{Einstein gravity as correct long-range theory}

\noindent
Einstein$\text{'}$s theory of gravitation is remarkably successful on long-distance scales, 
along with its low-energy Newtonian limit based upon the inverse-square law at short 
distances.  This has been verified over the range from binary pulsars to planetary 
orbits and short-distances on the order of 1 mm [134,135,38].  That conclusion is 
arrived at experimentally using spacecraft and lunar orbital measurements [136], as 
well as terrestrial laboratory tests of the inverse-square law  (ISL) [137,138], and 
the principle of equivalence [134].\\
\indent  
Similarly, Einstein gravity has prevailed experimentally over the JFBD scalar-tensor 
theory for the same distance scales.  Experimental limits on the JFBD parameter 
$\Omega$ from planetary time-delay measurements place it at best as $\Omega \geq 500$ 
while Cassini data indicates it may be $\Omega \geq 40,000$ [139].  For practical 
purposes, this is approximating the limit $\Omega \rightarrow \infty$ when one 
examines the PPN parameter $\gamma$ in solutions given in Appendix D, Case (a.2) 
where $\gamma \rightarrow 1$ in (D.8).  Further JFBD limits have been found in 
cosmology [139,Wu \& Chen; Weinberg].\\
\indent
This means that JFBD theory is basically Einstein gravity with $\gamma \rightarrow 1$.  
But these two theories are not equivalent, as shown here, because $\sigma$ is 
significantly related to $\lambda$ which in turn is the source of the CCP in Einstein 
gravity (\S 1.1).  In fact, it is $\sigma$ that helps solve the CCP.\\
\indent
In this context, no gravitational theory has been experimentally verified at the 
scales and energies that  are the focus of the present study, and now follow below.  
Hence, Einstein gravity ($\Lambda=0$) still prevails as the correct theory of gravity 
for all energies presently subject to experiment.  The present study does not change 
that well-established fact.

\subsection{Issues in and below the sub-millimeter regime}

\noindent
At short-distances scales below $1~cm$, the issue of what to measure is an entirely 
different matter.  There are virtually no experimental constraints on gravitational 
behavior at this range of interaction.\\
\indent
This scale eventually becomes the realm of hadron and high-energy particle physics.  
It is the realm that transitions from classical gravity to quantum gravity.  And it 
must address the physics of confinement $\textit{per se}$ because the graviton may play a 
pertinent role in that process.\\
\indent
\textit{Conventional Methods}.  The first experimental issue is the method of 
parameterization for identifying new forces and effects.  At the $mm$-scale, this is 
usually a comparison with a short-range Yukawa contribution to the familiar ISL 
$1/r^2$ term [137], as
\begin{equation}
V(r) = - V_o [1 + \alpha e^{-r/\lambda^\prime}]  
\end{equation}
\noindent
where $V_o$ is the ISL term, $\alpha$ is a dimensionless parameter, and 
$\lambda^\prime$ is a length scale or range.  The data are then published as graphs 
of $\mid \!\alpha \!\mid$ versus $\lambda^\prime$ [138].  It can be said that the ISL is valid 
down to $1~mm$ [137].\\
\indent
\textit{Limitations of Conventional Techniques}.  At these scales traditional 
experiments using torsion-balance or atomic microscopy techniques for studying the 
ISL, begin to encounter a strong background of nongravitational forces.  These 
include the Casimir and van der Waals forces [140].   Price [141] has pointed out 
that the experimentally accessible region for ISL study is limited to ranges greater 
than $40 ~\mu m$ by the electrostatic background force created by the surface potentials of 
metals and other materials.\\
\indent
\textit{Experimental Quantum Gravity (EQG)}.  Given that there is no consistent renormalizable theory of quantum gravity, there seems to have been little or no experimental work in quantum gravity at the short-distance scale.\\
\indent
In anticipation of the Large Hadron Collider (LHC) now operating at CERN, there has been much written about the onset of EQG at the TeV scale.  This includes dilatons and moduli from string theory, the leaking of gravitons into extra dimensions, M-theory, lowered Planck scales, and so forth.  Most are attempts to solve the SM hierarchy problem [137].

\subsection{Experimental consequences for scalar gluon mass}

\noindent
One of the first things to observe about the SM in particle physics is that it seems to 
ignore the scalar bag in hadron physics.  To see this, simply note that 
$\pounds_{SM} = \pounds_{EW} + \pounds_{QCD} + \pounds_{Int.}$ does not include 
any of the $\sigma$ terms in (2.7), (2.9), or (2.10)-(2.13).  If there is anything 
observable about $\sigma$ and the hadron bag, the SM is going to miss it.  Bag theory is 
apparently categorized as physics beyond the SM although no one seems to have 
pointed this out.\\
\indent
It is QCD that couples to the $\sigma$-field in (1.13) and (1.14).  Hence it is QCD that 
the present scalar-tensor theory must reckon with.  Since this study adopts the FLW 
NTS confinement model from the outset, its compatibility with QCD in the strong 
coupling regime has already been demonstrated [15,17,19,142].\\
\indent
What is new is the distinctive feature of the $\sigma$-field as a nonlinear, 
self-interacting scalar that represents the gluon condensate (or scalar gluon [18]) 
associated with hadron confinement (a bag), a broken symmetry in the QCD vacuum, the 
bag constant $B$, and Einstein$\text{'}$s cosmological constant $\lambda$.  This scalar 
$\sigma$-field has a classical mass $m_\sigma=\sqrt{a}$ in (B.38) and is a boson.  As 
mentioned in \S 1.3.1, it couples attractively to all hadronic matter in proportion to 
mass.  Hence, the $\sigma$-field has now become a fundamental field in scalar-tensor 
gravitation theory.\\
\indent
Note that the wave equation in (2.27) for $\sigma$ couples to the trace $T^M$ with 
mass contributions from the quark condensates $f \bar{\psi}\psi$ ($f\!\neq\!0$).  (2.27) 
states that the scalar gluon (condensate) $\sigma$ is observable as an exchange force.  
It makes predictions as to how $\sigma$ interacts with the quark condensates and all 
matter.\\
\indent
However, this does not mean that the bag is an observable in the laboratory.  Under 
high-temperature (\S 2.5) collisions, the bag can bifurcate or dissolve entirely 
(e.g., hadron decay).  The ultimate EQG question is whether the mass of the 
$\sigma$-field is a directly measurable quantity, much akin to measuring the gluon 
condensate in a free state which may include a quark-gluon plasma.  In another vein, 
the bag potential function $U(\sigma)$ or $U^*(\sigma)$ is not an observable.\\
\indent
So how can one determine $m_\sigma=\sqrt{a}$~?\\
\indent  
The mass $m_\sigma$ appears in all bag interaction potentials (2.19) either via SSB or 
when inserted by hand (such as Klein-Gordon or Pauli-Fierz).  In order to derive the 
mass $m_\sigma$ for comparison with experiment, it depends upon the parameterization 
of $U^*(\sigma)$ in (2.19).  There, the parameters $a,b,c$ (following the FLW NTS 
model) are interrelated and are used in conjunction with the bag constant $B$ to 
characterize a given hadron.  As an alternate choice, Creutz [64] uses 
$\alpha,\beta,\gamma$ and $B$.  To these, one adds $f$ in (2.12) and the strong 
coupling constant $\alpha_s=g_s/4\pi$ from (2.13).  In any case, based upon the 
characteristics defining confinement, one uses these parameters to construct 
$U^*(\sigma)$ and model the hadron at every level of approximation possible 
[19, pp. 21-22; 142], including temperature.\\
\indent
The answer, then, is that one takes the observed boson mass $m_\sigma$ and defines 
$\textit{a}$ in the confinement potential, as $a=m_\sigma^{~2}$.  That is one experimental 
observable that contributes to the definition of $U^*(\sigma)$, from which hadron dynamics 
(e.g. excited states) can be analyzed and predictions made.

\subsection{Experimental consequences for graviton mass inside the hadron}

\noindent
As for the subject of graviton mass, physics has yet to detect a graviton at all 
[134] – much less at the EQG scale of short-distance gravitation.  The EQG-scale 
confined graviton appears undetectable, much like the neutrino.   Hence its 
properties must be determined from things with which it interacts.  It also sheds most of 
its mass if a confined graviton emerges from the disintegrating hadron bag, shifting its 
mass $m_g$ from (2.32) to (2.31).\\
\indent
Nevertheless, inside the hadron the confined graviton acquires a mass via (2.32) and 
is shown in Table I.  As with all of the discussions of graviton physics at LHC 
energies mentioned above, an obvious thing to look for in exchange interactions is 
missing energy plus jets.  If graviton propagators are transporting energy and they 
cannot be detected, then this must show up as a missing energy.\\
\indent  
As an example, for the case of direct graviton production in say 
$p\bar{p} \rightarrow jet+graviton$, some have conjectured missing energy signatures 
[143].  However, for the graviton in Table I, the mass of a freed graviton is no 
longer (2.32) but rather (2.31) with a range that can reach a Hubble radius.  It is 
virtually massless at $\sim\!10^{-33}eV$.\\
\indent
In practice, it is difficult to tell experimentally the difference between quarks and 
gluons.  The reason is that both particles appear in the jets of hadrons [144].  
Confined graviton propagation may be even more difficult and much more tedious.\\
\indent
Finally, the graviton mass relation in (2.34) states that for a given vacuum energy 
density $B$ in the hadron interior ($B$=constant), the gravitational coupling 
constant $G$ does not have to be the Newtonian one in the exterior (\S 3.1.2).  
Since $G$ has never been experimentally determined in quantum gravity at sub-$mm$ 
scales, this is an important effect that needs to be addressed.  Changing 
$G_{Interior}$ moves the Planck scale in the interior.  If one moves the scale of 
the Planck mass ($M_{Pl} = 1/ \sqrt{G}$) , how is $G$ measured and determined?  One 
can consider this as a means for studying the SM hierarchy problem:  Increase $G$.  
However, how can it be proven experimentally?  The strategy is that the gravitational 
effects predicted in (2.26)-(2.27) can be made more significant by substantially increasing 
$G$, thus increasing the confined graviton mass (2.34) and spacetime curvature 
($R=4\lambda_{Bag}$) inside the hadron.  If those effects are experimentally 
established, then the issue is resolved.  Note that the excited state (the bag) at 
$<\sigma>=0$ does not change when modifying $G$ because $B$ is assumed (above) to be a 
constant.\\
\indent
If the Planck mass is moved significantly, the weak-field approximation of 
Appendix A is no longer valid.  The subject must then address the strong-field 
gravitational case which is beyond the scope of this study and quantum gravity as we 
understand it.

\subsection{Prospects in astrophysics and cosmology}

\noindent
A graviton mass has direct relevance to gravitational radiation research in 
astrophysics and cosmology [134,135].  This is important because gravitational wave 
astronomy is destined to become one of the new frontiers in our understanding of the 
Universe.\\
\indent
The bound on graviton Compton wavelength $m_g^{-1}$ derived for gravitational-wave 
observations of inspiralling compact binaries [135,38] is $m_g^{-1}\!\sim\!6 \cdot 
10^{17}~cm$.  From Table I, the Compton wavelength of a free graviton is 
$m_g^{-1}\!\sim\!10^{28}~cm$ which is eleven orders of magnitude safely beyond this 
experimental constraint.  A similar comparison applies to the velocity of graviton 
propagation.  Likewise, strong-field gravitational effects in stellar astrophysics 
(where Einstein gravity is known to prevail) are similarly unaffected by the small numbers in
Table I for the free graviton mass.

\subsection{Implications of Fermilab dilepton channel data about jets}

\noindent
The dilepton channel data observed at Fermilab [145] warrants comment from the point 
of view of hadron bag physics (\S 4.3 above).  During $p\bar{p}$ collisions at 
$\sim 2 ~TeV$, an unexpected peak has been found centered at $144 ~GeV/c^2$ during the 
production of a W boson which decays leptonically in association with two hadronic 
jets.\\
\indent
This could be a signal of a scalar gluon $\sigma$-field as discussed in \S 4.3 
during the production of jets.  If such a case proves plausible 
($m_\sigma=144 ~GeV/c^2$), then $a= m_\sigma^2= 2.07 \cdot 10^4 ~GeV^2/c^4$ in the 
hadron potential $U^*(\sigma)$ for the hadrons involved.\\
\indent  
However, it is well-known that annihilation energy (such as $e \bar{e}$ and 
$p \bar{p}$) can re-materialize into vector and scalar gluon jets [146].  Hence much 
additional work, involving the LHC, needs to address this subject.

\section{Comments and conclusions}

\subsection{Summary}

\noindent
A scalar-tensor theory of gravitation has been introduced as a modified 
Jordan-Fierz-Brans-Dicke model involving a scalar $\sigma$-field used in bag theory 
for hadron physics.  The two vacua (1.15) and (1.16), illustrated in Figure 1, have a 
natural explanation as a hadron inflated by a negative bag pressure $B$ in the 
gravitational ground-state background $\eta_{\mu \nu}$ of an accelerating 
Friedmann-Lemaitre (de Sitter) Universe.
\indent  
These results follow from having made the simple observation that the cosmological 
term $\lambda$ in Einstein gravity is a scalar potential function (\S B.1) and 
represents the confinement potential $U^*(\sigma)$ in $\pounds^*_{\sigma}$ 
(2.20) of hadron bag theory.  Since the $\sigma$-field in turn represents the gluon 
condensate (or gluon scalar [18]) in QCD as a scalar field, it is straight forward 
to conclude that scalar-tensor theory is a natural choice for introducing gravity –- 
albeit weak or strong - into particle physics at the TeV scale.\\
\indent
This scalar gluon couples to all hadronic matter uniformly, resulting in an 
attractive force proportional to hadron mass.  Hence it is a gravitational 
interaction.\\
\indent
Lee$\text{'}$s original motivation [18] for introducing the $\sigma$-field was to treat it as 
a phenomenological field that describes the collective long-range effects of QCD.  
There it has no short-wavelength components, so the $\sigma$-loop diagrams can be 
ignored leaving only tree diagrams.  That is, $\sigma$ has been regarded as a 
quasi-classical field.\\
\indent

\subsection{Postulates}

\noindent
Eight postulates or principal assumptions have been used, as follows:
\begin{enumerate}
  \item	The classical Einstein-Hilbert Lagrangian is augmented by a 
nonminimally coupled scalar NTS term $\pounds_{NTS}$ in the fashion of 
JFBD theory.  The $\pounds_{NTS}$ term represents hadron physics which 
includes QCD as $\pounds_{QCD}$ in the exact limit 
$\pounds_{NTS} \rightarrow \pounds_{QCD}$. 
  \item The gravitational field $g_{\mu \nu}$ couples minimally and universally to 
all of the fields of the Standard Model, as does Einstein gravity.  However, 
$g_{\mu \nu}$ also couples minimally and nonminimally to the composite features of 
hadron physics $\pounds_{NTS}$, not just $\pounds_{QCD}$.  This entails hadron physics.
  \item General covariance is necessary in order to define the procedure for the use 
of the Bianchi identities in determining conservation of energy-momentum from 
$T^M_{~~\mu \nu}$  in (2.5).  That means matter follows Einstein geodesics and obeys the 
principle of equivalence.  This assumption can be broken, applying the Bianchi 
identities to $T^*_{~\mu \nu}$ instead.  In such a case, the theory changes.  Also, 
use of the harmonic gauge (Appendix A) gives rise to a graviton mass, but breaks 
general covariance.
  \item Quantum vacuum fluctuations result in a broken vacuum symmetry, producing two 
distinct vacua containing two different vacuum energy densities $\lambda$.  Because 
$\lambda=\lambda(\mu,T)$, this broken symmetry is subject to restoration.
  \item The stability of the bag is assured by the vacuum energy density $B$ which is 
a negative vacuum pressure.
  \item The relation between graviton mass $m_g$ and $\lambda(\mu,T)$ found in the 
weak-field approximation, survives in the strong-field and strong-force cases.
  \item The NTS Lagrangian $\pounds_{NTS}$ is renormalizable.  The E-H 
Lagrangian $\pounds_{EH}$ is not, but it can be extended and made 
renormalizable at the sacrifice of unitarity.  Using the argument that unitarity is 
not required for the interiors of composite objects such as hadrons, this is less of 
a problem.  The expansion of the Lagrangian used here to include the additional terms 
of (B.1) then produces a tenable, renormalizable model for hadron physics that 
includes gravitation provided there is no chiral symmetry breaking ($d\!=\!0$).  
Deviation from unitarity, however, may signal the onset of new physics [147].
  \item The CCP has arisen because of inconsistent double-counting of $\lambda$ as a vacuum 
energy density both in QFT and in Einstein gravity - and with inconsistent dimensionality.  
\end{enumerate}

\subsection{Conclusions}

\noindent
Until now, a tensor theory with both short- and long-range gravitation has not been 
devised.  It has been shown in the weak-field approximation that this theory has both 
finite-range and short-range confined gravitational fields.  The short-range 
gravitational field is only present inside the hadron while gravity outside the 
hadron involves a free graviton possessing a tiny mass that reaches to the 
Hubble radius.  A guiding principle has been that gravity is universally coupled to 
all physics.  Hence that must include the composite features of hadron physics as 
well.\\
\indent  
What emerges is a conceivable confinement mechanism for the hadron bag that involves 
gravity.  In previous work, the bag was introduced $\textit{ad hoc}$ into flat-space 
using a Heaviside step function, or was explained with Lee$\text{'}$s color dielectric 
continuum model and Wilets' extension of it.  Here, however, gravitation is the 
origin of the vacuum energy density and is coupled directly to the scalar gluon 
(condensate) in QCD.\\
\indent
Finally, the study indicates that unification is another motivation for examining 
scalar-tensor theory in particle physics.  As shown in \S 4, these results are 
consistent with everything that is experimentally well established in QCD and 
Einstein gravity.

\section{Acknowledgments}

\noindent
The author would like to acknowledge communications with O.V. Pavlovsky and S.J. 
Aldersley.
\appendix
\section{The cosmological constant as a graviton mass}

\noindent
It was shown some time ago by this author [40] that the cosmological term $\lambda$ 
in General Relativity can be interpreted as a graviton mass, a result that will be 
reviewed in this Appendix.  Veltman subsequently [41] made a similar conclusion, 
except for Spin-2 quantum gravity - pointing out that the associated graviton 
propagators have negative probability.  However, Veltman$\text{'}$s result is not equivalent 
to what will be discussed here because he ``abandons from the start things like curved 
space.''  Spin-2 quantum gravity in flat space and quantized Einstein gravity are not 
the same thing since the latter is nonlinear and notoriously nonrenormalizable.  
Christensen \& Duff [42] have emphasized that quantizing Spin-2 gravity with 
$\lambda \neq 0$ must $\textit{not}$ be carried out by expansion in flat space, 
contrary to Veltman$\text{'}$s results.  One must consistently expand about a curved background 
field that satisfies the Einstein equations (1.1) with a $\lambda$ term.  There is also 
no compelling reason for singling out de Sitter space from the multitude of classical 
solutions where $\lambda \neq 0$.\\
\indent
The question, then, is to examine what is going on in General Relativity.  Is or is 
not $\lambda$ in (1.1) equivalent to or related to a graviton of non-zero rest mass in 
the sense of a wave equation?  Some say yes [43-45] while others say no [46-48] or 
declare that a graviton mass is impossible [49].  Note with caution that an 
unqualified graviton mass is beset with numerous problems in QFT.

\subsection{Weak-field limit, Kottler-Schwarzschild metric}

\noindent
The curved background adopted here will be a Kottler-Schwarzschild (KS) metric with 
$\lambda \neq 0$ [50] applied to the Regge-Wheeler-Zerilli (RWZ) problem [51] of 
gravitational radiation perturbations produced by a particle falling onto a large 
mass $M^*$.  The Einstein field equations (1.1) are repeated below for convenience:
\begin{equation}
R_{\mu \nu} - \frac{1}{2} g_{\mu \nu} R + \lambda g_{\mu \nu} = –- \kappa T_{\mu \nu}   \quad  . 
\end{equation}
\indent
One considers a small perturbative expansion of (A.1) about a known exact solution 
$\eta_{\mu \nu}$ subject to the boundary condition that $g_{\mu \nu}$  becomes 
$\eta_{\mu \nu}$ as $r \!\rightarrow\!\infty$.  The metric tensor $g_{\mu \nu}$ is thus 
assumed to be $g_{\mu \nu}$ = $\eta_{\mu \nu}$ + $h_{\mu \nu}$ where $h_{\mu \nu}$ is 
the dynamic perturbation such that $h_{\mu \nu} << \eta_{\mu \nu}$ = 
$g_{\mu \nu}^{(0)}$.  By virtue of Birkhoff$\text{'}$s theorem [52], the most 
general spherically symmetric solution is well-known to be a KS metric
\begin{equation}
ds^2 = -e^\nu dt^2 + e^\zeta dr^2 + r^2d\Omega^2  \quad ,
\end{equation}
\noindent
where 
\begin{equation}
e^\nu = 1 - \frac{2M}{r} - \frac{\lambda}{3}r^2 = e^{-\zeta}  \quad ,
\end{equation}
\noindent
while $M = GM^*/c^2, d\Omega^2=(d\theta^2 + \sin^2\theta d\phi^2)$, and 
$\eta_{\mu \nu}$ is determined from (A.1) as $\eta_{\mu \nu}=diag(-e^\nu, e^{-\nu}, 
r^2, r^2\sin2\theta )$ in spherically symmetric coordinates ($r,\theta,\phi$).  Its 
contravariant inverse is $\eta^{\mu \nu}$ defined as $\eta_{\mu \nu}\eta^{\mu \nu} 
= \delta_\mu^\nu$.  Note that when $M^*\!=\!0$ in (A.1) a de Sitter space results and 
photons following geodesics do not travel at the speed of light $c$.  Hence 
$\lambda \neq 0$ implies a photon rest-mass [40,53].  Although this geometric 
property of curved backgrounds has been often ignored by gauge theorists, it does 
not mean disaster for gauge invariance.  Goldhaber \& Nieto [54] have provided a very 
nice discussion of the fact that Stueckelberg$\text{'}$s construction [55] removes the 
formal gauge-invariance argument for a zero photon mass (and certainly for curved 
backgrounds).  Gauge invariance does not forbid an explicit mass term for the gauge 
field should the graviton be the gauge boson.\\
\indent
The wave equation for gravitational radiation $h_{\mu \nu}$ on the non-flat 
background containing $\lambda$ in (A.1) follows as (A.20) below, derived now from the 
formalism developed for studying the RWZ problem.  Perturbation analysis of (A.1) for a stable 
background $\eta_{\mu \nu} = g_{\mu \nu}^{(0)}$ produces the following
\begin{equation}
[h_{\mu\nu;\alpha}^{\quad~;\alpha} - h_{\mu\alpha;\nu}^{\quad~;\alpha} - h_{\nu\alpha;\mu}^{~\quad;\alpha} + h_{\alpha~;\mu;\nu}^{~\alpha}] + \eta_{\mu\nu}[h_{\alpha\gamma}^{~~;\alpha;\gamma} - h_{\alpha ~;\gamma}^{~\alpha ~;\gamma}] + h_{\mu\nu}(R - 2\lambda) - \eta_{\mu\nu}h_{\alpha\beta}R^{\alpha\beta}  = - 2\kappa ~\delta T_{\mu \nu}.
\end{equation}
\noindent
Stability must be assumed in order that $\delta T^{\mu \nu}$ is small.  This equation 
can be simplified by defining the function (introduced by Einstein himself) 
\begin{equation}
\bar{h}_{\mu\nu} \equiv h_{\mu\nu} - \frac{1}{2}\eta_{\mu\nu} h
\end{equation}
and its divergence
\begin{equation}
f_\mu \equiv \bar{h}_{\mu\nu}^{~~ ;\nu} \quad	.
\end{equation}
\indent
Substituting (A.5) and (A.6) into (A.4) and re-grouping terms gives
\begin{equation}						
\bar{h}_{\mu\nu;\alpha}^{\quad~;\alpha} - (f_{\mu;\nu} + f_{\nu;\mu}) + \eta_{\mu\nu}f_{\alpha}^{~;\alpha} - 2\bar{h}_{\alpha\beta}R^{\alpha ~~ \beta}_{~\mu\nu} - \bar{h}_{\mu\alpha}R^\alpha_{~\nu} - \bar{h}_{\nu \alpha}R^\alpha_{~\mu}
\end{equation}
\noindent
\begin{equation}						
 + h_{\mu\nu}(R - 2\lambda)- \eta_{\mu\nu}h_{\alpha\beta}R^{\alpha\beta} = - 2\kappa \delta T_{\mu\nu} \nonumber  \quad .
\end{equation}
Now impose the Hilbert-Einstein-de Donder gauge which sets (A.6) to zero
\begin{equation}
f_\mu = 0  \quad  ,
\end{equation}
\noindent
and suppresses the vector gravitons.  ($f_\mu \neq 0$ can be retained for further 
simplification in some cases of $\eta_{\mu \nu}$, although problematic negative 
energy states may be associated with these vector degrees of freedom.)  (A.8) now 
reduces wave equation (A.7) to
\begin{equation}
\bar{h}_{\mu\nu;\alpha}^{\quad~;\alpha} - 2\bar{h}_{\alpha\beta}R^{\alpha ~~ \beta}_{~\mu\nu} - \bar{h}_{\mu\alpha}R^\alpha_{~\nu} - \bar{h}_{\nu \alpha}R^\alpha_{~\mu} - \eta_{\mu\nu}h_{\alpha\beta}R^{\alpha\beta} + h_{\mu\nu}(R - 2\lambda) = -2\kappa ~\delta T_{\mu\nu}  \quad  .
\end{equation}
\noindent
In an empty ($T_{\mu\nu}=0$), Ricci-flat ($R_{\mu\nu}=0$) space without $\lambda$ 
($R=4\lambda=0$), (A.9) further reduces to 
\begin{equation}
\bar{h}_{\mu\nu;\alpha}^{\quad~;\alpha} - 2R^{\alpha ~~ \beta}_{~\mu\nu}\bar{h}_{\alpha\beta} = -2\kappa ~\delta T_{\mu\nu}   ,
\end{equation}
\noindent
which is the starting point for the RWZ formalism.

\subsection{Weak-field limit, de Sitter metric}

\noindent
Since $\lambda \neq 0$ is of paramount interest here, we know that the trace of the field 
equations (A.1) gives
\begin{equation}
4\lambda - R = -\kappa T   
\end{equation}
\noindent
whereby they become
\begin{equation}
R_{\mu \nu} - \lambda g_{\mu \nu} = –\kappa [T_{\mu \nu} - \frac{1}{2}g_{\mu \nu} T]  \quad .
\end{equation}
\noindent
For an empty space ($T_{\mu\nu} = 0$ and $T=0$), (A.12) reduces to de Sitter space
\begin{equation}
R_{\mu \nu} = \lambda g_{\mu \nu}  \quad 
\end{equation}
\noindent
and the trace (A.11) to 
\begin{equation}
R = 4\lambda  \quad . 
\end{equation}
\indent
Substitution of (A.13) and (A.14) into (A.9) using (A.5) shows that the contributions due 
to $\lambda \neq 0$ are now of second order in $h_{\mu\nu}$.  Neglecting these terms 
(particularly if $\lambda$ is very, very small) simplifies (A.9) to
\begin{equation}
\bar{h}_{\mu\nu;\alpha}^{\quad~;\alpha} - 2R^{\alpha ~~ \beta}_{~\mu\nu}\bar{h}_{\alpha\beta} = -2\kappa ~\delta T_{\mu\nu}  \quad .
\end{equation}
\noindent
Note that one can arrive at (A.15) to first order in $h_{\mu\nu}$ by using 
$g_{\mu\nu}$ as a raising and lowering operator rather than the background 
$\eta_{\mu\nu}$ –- a result which incorrectly leads some [48] to the conclusion that 
$\lambda$ terms cancel in the gravitational wave equation.\\
\indent
Furthermore, note with caution that (A.15) and the RWZ equation (A.10) are \textit{not} 
the same wave equation.  Overtly, the cosmological terms have vanished from (A.15), 
just like (A.10) where $\lambda$ was assumed in the RWZ problem to be nonexistent in 
the first place.  However, the character of the Riemann tensor $R^{\alpha\mu\nu\beta}$ 
is significantly different in these two relations.\\
\indent  
Let us simplify the KS metric by setting the central mass $M^*$ in $\eta_{\mu \nu}$  
to zero.  This produces the de Sitter space (A.13)-(A.14) of constant curvature $K = 1/R^2$, 
where we can focus on the effect of $\lambda$.  The Riemann tensor is now
\begin{equation}
R_{\gamma\mu\nu\delta} = +K (g_{\gamma \nu} g_{\mu \delta} - g_{\gamma \delta}g_{\mu \nu}) 
\end{equation}
\noindent
and reverts to
\begin{equation}
R^{\alpha~~\beta}_{~\mu\nu} = +K (g^{\alpha}_{~\nu}g_{\mu}^{~\beta} - 
g^{\alpha \beta}g_{\mu \nu}) 
\end{equation}
\noindent
for use in (A.15).  This substitution (raising and lowering with $\eta_{\mu \nu}$) 
into (A.15) now gives a $K$ and a $\lambda$ term contribution 
\begin{equation}
-2K[(\bar{h}_{\mu\nu} - \eta_{\mu\nu}\bar{h}) + (\bar{h}_{\alpha\mu}h^\alpha_{~\nu} + \bar{h}_{\nu\beta}h^\beta_{~\mu} - \bar{h}h_{\mu\nu} - \eta_{\mu\nu}h^{\alpha\beta}\bar{h}_{\alpha\beta})] + \lambda[2\bar{h}_{\mu\alpha}h^\alpha_{~\nu} + \eta_{\mu\nu}h_{\alpha\beta}^2 ]
\end{equation}
\noindent
to second order in $h_{\mu \nu}$.  Recalling that curvature $K$ is related to 
$\lambda$ by $K = \lambda/3$, substitution of (A.18) back into (A.15) gives to first 
order
\begin{equation}
\bar{h}_{\mu\nu;\alpha}^{\quad~;\alpha} - \frac{2}{3}\lambda\bar{h}_{\mu\nu} + \frac{2}{3}\lambda\eta_{\mu\nu}\bar{h} = -2\kappa ~\delta T_{\mu\nu}  \quad . 
\end{equation}
\noindent
There is no cancellation of the $\lambda$ contributions to first order.  Noting from 
(A.5) that $\bar{h} = h(1- \frac{1}{2}\eta)$, then a traceless gauge $\bar{h} = 0$ 
means either that $h = 0$ or $\eta = 2$.  Since $\eta = 4$, (A.19) reduces to
\begin{equation}
\bar{h}_{\mu\nu;\alpha}^{\quad~;\alpha} - \frac{2}{3}\lambda\bar{h}_{\mu\nu} = -2\kappa ~\delta T_{\mu\nu} 
\end{equation}
\noindent
in a traceless Hilbert-Einstein-de Donder gauge where $\bar{h}_{\mu\nu}^{~~;\nu}=\!0$ 
and $\bar{h}_\mu^{~\mu}=\!0$.  (A.20) is a wave equation involving the Lapace-Beltrami 
operator term $\bar{h}_{\mu\nu;\alpha}^{~\quad ;\alpha}$ for the Spin-2 gravitational 
perturbation $\bar{h}_{\mu\nu}$ bearing a mass
\begin{equation}
m_g = \sqrt{2\lambda/3}  \quad , 
\end{equation}
\noindent
similar to the Klein-Gordon equation  $(\square –- m^2)\phi = 0$ for a Spin-0 scalar 
field $\phi$ in flat space.\\  
\indent
\textit{Summary.}  Because the trace $T$ was assumed to vanish in step (A.13)-(A.14), the 
scalar graviton (since it couples to $T\!\neq\!0$) has been suppressed along with the 
two vector Spin-1 components by virtue of the gauge condition $f_\mu\!=\!0$, leaving 
only two transverse degrees of freedom of the $(2S\!+\!1)\!=\!5$ helicities for a 
massive graviton.  One can study further expansions of (A.20) to show that the 
$\lambda$ term survives but this has been done elsewhere [40].\\
\indent
Hence a well-behaved massive graviton containing two transverse degrees of freedom 
has been obtained without ghosts (preserving unitarity) in the weak-field 
approximation.  This has been accomplished by not introducing the traditional 
Pauli-Fierz [56] mass term 
$\mathfrak{L}_{PF}=\frac{1}{4}m_{PF}^2(h_{\mu\nu}h^{\mu\nu}\!- \!h^{\mu ~2}_{~\mu})$ 
which is often described as the only ghost-free form for a Spin-2 particle [52].  Spin-2 
graviton ghost problems can also be averted by beginning with a scalar-tensor 
theory [13, Duff] as is done in this study.\\
\indent
Note that general covariance has been broken by going to the 
Hilbert-Einstein-De Donder gauge $f_\mu \!=\!0$ in (A.8) in order to suppress the 
vector gravitons.  Note also that radiation reaction is a direct dividend of the 
nonlinear Einstein theory which is not accounted for in the linearization used by 
the RWZ- or KS-formalism employed here.\\
\indent
Finally, Duff $\textit{et al.}$ [57] have corroborated the results presented here 
that $m_g\!=\!\sqrt{2\lambda/3}$ in (2.31)-(2.32), as well as Higuchi [58] save for a 
factor of two.

\subsection{Conformal invariance and mass}
\noindent
As a sanity check, consider the following.  Penrose [59] showed that the zero rest-mass 
free-field equations for each spin value are conformally invariant if interpreted suitably.  
For a massless Spin-0 field $\phi$ on a background with scalar curvature $R$, the wave 
equation is
\begin{equation}
(\square - \frac{R}{6})\phi  \quad , 
\end{equation}
\noindent
a result that can be generalized to arbitrary integer spin [60].  According to the 
Klein-Gordon equation $(\square –- m^2)\phi\!=\!0$ for such a field in flat space, 
one concludes that $\phi$ has a mass $\!m=\!\sqrt{R/6})$.  For the graviton case 
(A.20) in the previous section, substituting (A.14) or $R\!=\!4\lambda$ into (A.22) 
represents a de Sitter space whereby 
\begin{equation}
(\square - \frac{2}{3}\lambda)\phi = 0  \quad . 
\end{equation}
\noindent
We have recovered precisely the weak-field wave equation (A.20), except as the Spin-0 
component of a graviton with the same mass $m_g\!=\!\sqrt{2\lambda/3}$.

\section{Scalar-tensor theory with Jordan-Fierz-Brans-Dicke \textit{ansatz}}

\subsection{Fundamentals of scalar-vector-tensor theory}

\noindent
The prevailing theory of gravitation is Einstein gravity (\S 4) whose Lagrangian (2.1) 
provides the field equations (1.1).  It is a scalar-vector-tensor theory in which its 
field tensor $g_{\mu \nu}$ consists of sixteen independent variables that interact 
through the energy-momentum tensor $T_{\mu \nu}$ with other fields.  $T_{\mu \nu}$ 
is comprised of a Spin-2 (tensor), three Spin-1 (vectors), and two Spin-0 (scalars) 
admixtures totaling 16 degrees of freedom or helicities.  By assuming symmetry 
$T_{\mu \nu}=T_{\nu \mu}$ two of the Spin-1 (vector) admixtures are suppressed.  
Energy conservation $T_{\mu \nu}^{~~;\nu}=0$ eliminates the remaining vector and one 
of the scalars.  The final scalar can be removed by ensuring that there is no trace 
($T_{\mu}^{~\mu}\!=\!0$) which can interact with $g_{\mu \nu}$.  The result is a 
well-behaved, consistent, massless graviton in its quantum gravity version.\\
\indent
The Lagrangian is used to bring the above dynamics together.  Ideally, there might 
be one scalar field $\phi$, one vector field $A_\mu$ such as Yang-Mills or Maxwell or 
both, and the graviton that work together in a unified fashion.  Since the focus 
here is on the scalar field contribution in curved backgrounds, we can discuss a 
generic Lagrangian using three simple scalar densities:  
$\sqrt{-g}R$, $\sqrt{-g}\mathfrak{B}$, and $\sqrt{-g}$ where $\mathfrak{B}$ 
represents any of the Lorentz scalar interactions allowable under the inhomogeneous 
group discussed in Appendix C.2, although many of these can be introduced by simply 
re-defining the covariant derivative $\nabla_\mu$ in the sense of gauge invariance.  
Noting that there must also be a kinematic term for the gradient of the scalar field 
$\phi$, an example of such a general Lagrangian in four dimensions is as follows:
\begin{equation}
\pounds = \sqrt{-g}[f_1(\phi)R + f_2(\phi)\mathfrak{B} + f_3\nabla_{\mu}\phi\nabla^{\mu}\phi - \lambda(\phi)]  \quad , 
\end{equation}
\noindent
recognizing that $\phi(\lambda)\sqrt{-g}$  is the cosmological term and is a function 
of the scalar field $\phi$.  It actually is a scalar potential function 
$\lambda(\phi)=U(\phi)$ which determines the vacuum energy density.  Since Lagrangians 
$\pounds=T-U$ are kinetic energy ($T$) minus potential energy ($U$), (B.1) 
can be also written
\begin{equation}
\pounds = \sqrt{-g}[f_1(\phi)R + f_2(\phi)\mathfrak{B} + f_3\nabla_{\mu}\phi\nabla^{\mu}\phi - U(\phi)]  \quad . 
\end{equation}
\noindent
To the right-hand-side must be added the source term for matter 
$\pounds_{matter}$ that produces $T_{\mu \nu}$ .  This discussion is the 
general idea for the scalar portion of the scalar-tensor theory and what follows in 
Appendix B.  One can see that the nonlinear $\sigma$-field Lagrangian 
$\pounds_\sigma$ for the bag in (2.11) and (2.20) appears naturally in the 
right-hand-side of (B.2).

\subsection{Basic derivations in support of field equations (2.26)-(2.27)}

\noindent
In its original form, JFBD theory [67,68] did not include a potential $U(\sigma)$ 
or $\lambda$.  The E-H action (2.1) was used with $\lambda\!=\!0$.  Since the theme of 
the present study is $\lambda$ with major emphasis on the $\lambda\!=\!\lambda(\phi)$ 
term in (B.1), although with the substitution $\phi\!\rightarrow\!\sigma$ 
representing the scalar $\sigma$-field, the scalar-tensor theory must be modified 
(denoted as $\lambda$JFBD).\\
\indent
JFBD made the assumption that the reciprocal of Newton$\text{'}$s gravitational coupling 
constant $G^{-1}$ is to be replaced by a scalar field $\sigma$.  This is known as 
the JFBD \textit{ansatz}:
\begin{equation}
\kappa = \sigma^{-1}  
\end{equation}
\noindent
which is adopted here because this is how scalar-tensor theory began.\footnote{Use of 
$\phi^2 R$ for nonminimal coupling [62] has an advantage when considering Higgs gravity 
because it manifestly represents $\phi^2 \rightarrow \Phi^\dag\Phi$ as the Higgs doublet.  
However, this disguises the results here which use (B.3).}  It needs to be noted that 
$\kappa\!\rightarrow\!\kappa(\phi)$ can be any permissible function of the scalar field 
provided this results in consistent physics.  A non-permissible example would a polynomial 
of degree $n\!=\!5$ including $\sigma^{5}$ in (1.11) which by dimensional counting would 
result in dimensional coefficients that produce a nonrenormalizable potential $U^*(\sigma)$.  
See \S B.3 for more.\\
\indent
The $\lambda$JFBD Lagrangian $\pounds_{\lambda JFBD}$, assuming (B.3) and including a kinetic term for 
$\sigma$ while re-instating potential $U^*(\sigma)$ and $\lambda$, is then
\begin{equation}
\pounds_{\lambda JFBD} = \frac{1}{2}\sqrt{-g}[-\sigma R + \frac{\Omega}{\sigma}\nabla_{\mu}\sigma\nabla^{\mu}\sigma - U^*(\sigma)] +8\pi\pounds_{matter}  
\end{equation}
\noindent
where $\Omega$ is the dimensionless JFBD coupling constant.  (B.4) is basically an 
extension of the Jordan-type action in [67].  Once again, it has been modified by the 
presence of a vacuum energy density potential $U^*(\sigma)$.  Variation $\delta S=0$ 
of (2.3) using (B.4) gives the field equations for $g_{\mu \nu}$ and $\sigma$ to be 
derived below.\\
\indent
The energy-momentum tensor in (2.5) is comprised of two terms: $T^*_{~\mu \nu} =  
T^M_{~\mu \nu} + T^{\sigma}_{~\mu \nu}$.  First is  the usual matter contribution 
$T^M_{~\mu \nu}$
\begin{equation}						
T^M_{~~\mu \nu} = \frac{2}{\sqrt{-g}}[\frac{\partial(\sqrt{-g}\pounds_M)}{\partial g^{\mu \nu}} - \partial^\alpha\frac{\partial(\sqrt{-g}\pounds_M)}{\partial(\partial^\alpha g^{\mu \nu})} ]
\end{equation}
\noindent
which includes all matter fields in the Universe except gravitation, and it is 
assumed to be independent of the $\sigma$-field.\\
\indent  
Characteristic of the JFBD theory, there is a new term $T^{\sigma}_{~\mu \nu} = 
\nabla_{\mu}\sigma\nabla_{\nu}\sigma  -  g_{\mu \nu}\pounds^*_{\sigma}$ which must 
include the effects of $\pounds_{G,\sigma}$ in (2.18).  Consolidating all of the 
$\sigma$ terms and introducing a superscript "R" for renormalizable, we have in 
short-hand derivative notation
\begin{equation}						
^RT^{\sigma}_{~\mu \nu} = \sigma_{;\mu}\sigma_{;\nu} - \frac{1}{2}g_{\mu \nu}\sigma_;^\alpha\sigma_{;\alpha} + g_{\mu \nu}U^*(\sigma)  \quad . 
\end{equation}
\noindent
With (B.5) and (B.6), variation of (B.4) will give the final result (2.26)-(2.27) in the 
text as shown below.\\
\indent  
A principal assumption follows Brans and Dicke.  One does not want to sacrifice the 
success of the principle of equivalence in Einstein$\text{'}$s theory [38].  Hence only 
$g_{\mu \nu}$ and not $\sigma$ enters the equations of motion for matter consisting 
of particles and photons.  The interchange of energy between matter and gravitation 
thus must follow geodesics as assumed by Einstein [69].  Therefore, the 
energy-momentum tensor (B.5) is assumed to be conserved in the standard fashion, 
$T^{M ~~;\nu}_{~~\mu \nu}\!=\!0$ (for exceptions see Footntes (9) and (11), as well as [70]).  
This also places an important constraint on the Spin-2 degrees of freedom in the 
quantized version.\\
\indent
Now it is time to focus on $T^{\sigma}_{~\mu \nu}$ in (B.6). The most general 
symmetric tensor of the form (B.6) which can be built up from terms each of which 
involves two derivatives of one or two scalar $\sigma$-fields, and $\sigma$ itself, 
is
\begin{equation}						
T^{\sigma}_{~\mu \nu} = A(\sigma)\sigma_{;\mu}\sigma_{;\nu} + B(\sigma)\delta_{\mu\nu}\sigma_{;\alpha}\sigma^{;\alpha} + C(\sigma)\sigma_{;\mu;\nu} + D(\sigma)\delta_{\mu\nu}\square\sigma + E(\sigma)g_{\mu \nu}U^*(\sigma)  \quad . 
\end{equation}
\indent
We want to find the coefficients $A, B, C, D,$ and $E$.  One can make the argument 
that the last term in (B.7) \underline{is} $g_{\mu \nu}U^*(\sigma)$, whereby 
$E(\sigma)\equiv 1$ (\S B.3), but we will carry $E(\sigma)$ along at the present time.\\
\indent
Recalling that
\begin{equation}						
\nabla_\mu U(\sigma) = \frac{dU^*}{d\sigma}\frac{d\sigma}{d\mu} \equiv U^{*\prime}(\sigma)\sigma_{;\mu} \quad , 
\end{equation}
\noindent
the covariant divergence of (B.6) is
\begin{equation}						
^RT^{\sigma~\mu}_{\quad~\nu ;\mu} = \sigma_{;\nu}\square \sigma -\sigma_{;~;\nu}^\mu \sigma_{;\mu} + U^{*\prime}(\sigma)\sigma_{;\nu}  \quad , 
\end{equation}
\noindent
and the covariant divergence of (B.7) is 
{\setlength\arraycolsep{2pt}
\begin{eqnarray}
T^{\sigma~\mu}_{~\quad\nu ;\mu} & ~= &  \quad[A(\sigma) + B^\prime(\sigma)]\sigma_{;}^\mu \sigma_{;\nu}\sigma_{;\mu} {}
					\nonumber\\
	& & {} + [A(\sigma) + D^\prime(\sigma)]\sigma_{;\nu}\square \sigma  {}
					\nonumber\\
	& & {} + [A(\sigma) + 2B(\sigma) +  C^\prime(\sigma)]\sigma_{;~;\nu}^\mu \sigma_{;\mu}  {}
					\nonumber\\
	& & {} + [D(\sigma) ](\square \sigma)_{;\nu}  {} 
					\nonumber\\
	& & {} + [C(\sigma) ]\square (\sigma)_{;\nu})  {}
					\nonumber\\
	& & {} + [E(\sigma)U^{*\prime}(\sigma) + U^*(\sigma)E^{\prime}(\sigma) ]\sigma_{;\nu}  \quad .
\end{eqnarray}}\\
\indent
Multiplying the field equations (2.4) and (2.5) by $\sigma$, one obtains
\begin{equation}						
(R_{\mu \nu} - \frac{1}{2}g_{\mu \nu}R)\sigma = - 8\pi T^M_{~~\mu \nu} - 8\pi T^{\sigma}_{~\mu \nu}  \quad . 
\end{equation}
\noindent
Taking the divergence of (B.11) gives
\begin{equation}						
(R_{\mu \nu} - \frac{1}{2}g_{\mu \nu}R);^\mu \sigma + (R_{\mu \nu} - \frac{1}{2}g_{\mu \nu}R)\sigma_;^\mu = - 8\pi T^{M ~~\mu}_{~\mu \nu;}  - 8\pi T^{\sigma ~~\mu}_{~\mu \nu;}  \quad . 
\end{equation}
\noindent
The first term on the left-hand-side of (B.12) is zero by virtue of the Bianchi 
identities.  The first term on the right-hand-side is zero because 
$T^{M ~~;\nu}_{~~\mu \nu}\!=\!0$ and is conserved in order that matter follows 
Einstein geodesics (principle of equivalence).  Next turning to an identity 
involving the Riemann tensor $R^\gamma_{~\alpha\nu\beta}$, first and second 
derivatives of a covariant vector $A^\gamma$ contain an antisymmetric part [69]
\begin{equation}						
A^\gamma_{~;\nu;\beta} - A^\gamma_{~;\beta;\nu} = A^\alpha R^\gamma_{~\alpha \nu\beta}  \quad . 
\end{equation}
\noindent
This relation (B.13) means the first non-zero term in (B.12) is 
\begin{equation}						
R_{\mu \nu}\sigma_;^\mu = \sigma^{~\alpha}_{;~;\alpha;\nu} - \sigma^{~~\alpha}_{;\nu; ~;\alpha} = (\square \sigma)_{;\nu} - \square (\sigma_{;\nu})  \quad . 
\end{equation}
\noindent
Taking the trace of (2.4) and (2.5)
\begin{equation}						
R = \kappa T^M + \kappa T^\sigma  \quad  
\end{equation}
\noindent
and using the equation of motion for $\sigma$ (with $\sigma$-quark coupling constant $f=0$) to include the gravitational coupling to the 
trace $T^M$ of Einstein gravity, 
\begin{equation}						
\square\sigma = \frac{1}{2}\kappa_1 T^M + U^{*\prime}(\sigma)  \quad  
\end{equation}
\begin{equation}						
T^M = 2\kappa_1^{-1} (\square\sigma - U^{*\prime}(\sigma))  \quad . 
\end{equation}
\noindent
From (B.7) we obtain the other trace 
\begin{equation}						
T^\sigma = [A(\sigma) + 4B(\sigma)]\sigma^{;\alpha} \sigma_{;\alpha} + [C(\sigma)+4D(\sigma)]\square\sigma + 4[E(\sigma)U^*(\sigma)]  \quad . 
\end{equation}
\noindent
It follows from (B.15), (B.16), and (B.18) that 
\begin{equation}						
R = 2\kappa_1^{-1}(\square\sigma + U^{*\prime}(\sigma)) + \kappa([A(\sigma) + 4B(\sigma)]\sigma^{;\alpha}\sigma_{;\alpha} + [C(\sigma) + 4D(\sigma)]\square\sigma + 4[E(\sigma)U^*(\sigma)])  \quad . 
\end{equation}
\noindent
The left-hand side of (B.12), using (B.13) and (B.19), becomes
\begin{equation}						
(R_{\mu \nu} - \frac{1}{2}g_{\mu \nu}R)\sigma_;^\mu  =  (\square \sigma)_{;\nu} - \square (\sigma_{;\nu}) - \frac{1}{2}\big[\kappa \kappa_1^{-1}(\square\sigma + U^{*\prime}(\sigma)) \nonumber
\end{equation}
\noindent
\begin{equation}						
\qquad \quad+ \kappa\big([A(\sigma) + 4B(\sigma)]\sigma^{;\alpha}\sigma_{;\alpha} + [C(\sigma)+4D(\sigma)]\square \sigma + 4E(\sigma)U^*(\sigma)\big)\big]   ~~~ .
\end{equation}
\noindent
Now rearrange (B.20) for comparison with (B.10):\\
{\setlength\arraycolsep{2pt}
\begin{eqnarray}						
(R_{\mu \nu} - \frac{1}{2}g_{\mu \nu}R)\sigma_;^\mu & = & -\frac{1}{2}\kappa[A^\prime(\sigma) + 4B(\sigma)]\sigma_;^{~\mu} {}
					\nonumber\\
	& & {} -\frac{1}{2}\kappa[2\kappa_1^{-1}+C(\sigma)+D(\sigma)]\sigma_{;\nu}\square \sigma  {}
					\nonumber\\
	& & {} + [0]\sigma_{;~~;\nu}^{~\mu}\sigma_{;\mu}  {}
					\nonumber\\
	& & {} + [1](\square \sigma)_{;\nu}  {} 
					\nonumber\\
	& & {} + [-1]\square (\sigma_{;nu})  {}
					\nonumber\\
	& & {} -\frac{1}{2}\kappa[2\kappa_1^{-1}U^{*\prime}(\sigma)+4E(\sigma)U^*(\sigma)]\sigma_{;\nu}   \quad. 
\end{eqnarray}}
\indent
In order that (B.12) be true, substituting (B.9) and (B.21), the bracketted 
coefficients in (B.10) and (B.21) must be equal term by term.  This requires the 
following:
\begin{equation}						
1 = -8\pi D(\sigma) 
\end{equation}
\noindent
\begin{equation}						
-1 = -8\pi C(\sigma) 
\end{equation}
\noindent
\begin{equation}						
\frac{1}{2}\kappa[\kappa_1^{-1} + C(\sigma) +4D(\sigma)] = 8\pi[A(\sigma) + D^\prime (\sigma)]
\end{equation}
\noindent
\begin{equation}						
\frac{1}{2}\kappa[A(\sigma)+4B(\sigma)]= -8\pi[A^\prime(\sigma)+B^\prime(\sigma)] 
\end{equation}
\noindent
\begin{equation}						
0 = A(\sigma) + 2B(\sigma) + C^\prime(\sigma) 
\end{equation}
\noindent
\begin{equation}						
\frac{1}{2}\kappa[2\kappa_1^{-1}[U^{*\prime}(\sigma) + 4E(\sigma)U^*(\sigma)] = 8\pi[E(\sigma)U^{*\prime}(\sigma)+U^*(\sigma)E^\prime(\sigma)]
\end{equation}
\indent
Let us find the solution of (B.22)-(B.27), determining $A, B, C,$ and $D$.  Then we will 
address $E(\sigma)$ in (B.27) in Appendix B.3.  From (B.22) and (B.23)
\begin{equation}						
C(\sigma)=-D(\sigma)= - 1/8 \pi 
\end{equation}
\noindent
\begin{equation}						
C^\prime(\sigma)=D^\prime(\sigma)=0  \quad . 
\end{equation}
\noindent
From (B.24), one has $A(\sigma) = \frac{1}{2}\kappa[\kappa_1^{-1} - \frac{3}{2}]$.  
Define
\begin{equation}						
\Omega = \kappa_1^{-1} - \frac{3}{2}  \quad , 
\end{equation}
\noindent
whereby $\kappa_1$ in (B.16) and (B.17) is
\begin{equation}						
\kappa_1 = \frac{2}{3+2\Omega}  \quad , 
\end{equation}
\noindent
Then
\begin{equation}						
A(\sigma) = \frac{\Omega}{8\pi \sigma}  \quad . 
\end{equation}
\noindent
Using (B.23) and (B.26) gives
\begin{equation}						
B(\sigma) = - \frac{1}{2}A(\sigma)  \quad . 
\end{equation}
\noindent
\begin{equation}						
B^\prime(\sigma) = - \frac{1}{2}A^\prime(\sigma)  \quad . 
\end{equation}
\noindent
Substitution of (B.32), (B.33), (B.22), and (B.23) into (B.7) results in
\begin{equation}						
\kappa T^{\sigma}_{\mu \nu} = \frac{\Omega}{\sigma^2}[\sigma_{;\mu}\sigma_{;\nu} - \frac{1}{2}g_{\mu \nu}\sigma_{;\alpha}\sigma^{;\alpha}] - \frac{1}{\sigma}[\sigma_{;\mu}\sigma_{;\nu} - g_{\mu \nu}\square\sigma] - \frac{1}{\sigma}[E(\sigma)g_{\mu \nu}U^*(\sigma)]  \quad . 
\end{equation}
\indent
Inserting (B.35) into (2.4) and (2.5) of the text gives the full field equation
\begin{equation}						
(R_{\mu \nu} - \frac{1}{2} g_{\mu \nu} R) = - \frac{8\pi}{\sigma}T^M_{\mu \nu} - 
\frac{\Omega}{\sigma^2}[\sigma_{;\mu}\sigma_{;\nu} - 
\frac{1}{2}g_{\mu \nu}\sigma_{;\alpha}\sigma^{;\alpha}] - 
\frac{1}{\sigma}[\sigma_{;\mu}\sigma_{;\nu} - g_{\mu \nu}\square \sigma] - 
\frac{1}{\sigma}[E(\sigma)g_{\mu \nu}U^*(\sigma)]  \quad , 
\end{equation}
\noindent
while (B.31) in (B.16) gives the scalar wave equation (for $f\!=\!0$)
\begin{equation}						
\square \sigma = \frac{8\pi}{3+2\Omega}T^* + U^{*\prime} (\sigma)  \quad , 
\end{equation}
\noindent
provided $\Omega$ cannot be equal to $-3/2$.  If so, (B.36) is a conformally mapped 
set of Einstein field equations.  For $f\!\neq \!0$, (B.36) and (B.37) are simply (2.26) and 
(2.27) of the text and are the field equations for this scalar-tensor theory.  $E(\sigma)$ is 
examined in \S B.3 that follows.\\
\indent
From the dimensionality of $U^*(\sigma)$ in (2.19), we see that $a$ has 
mass-dimension two or $m^2$.  Taking its derivative $U^{*\prime}(s)$ in conjunction 
with (2.24) and (2.27), the $\sigma$–-field has a mass
\begin{equation}						
m_\sigma = \sqrt{a}  \quad . 
\end{equation}
\indent
Hence it is a short-range field.

\subsection{Discussion of auxiliary equation (B.27)}

\noindent
The purpose for having introduced $E(\sigma)$ has been to conform with the criteria 
for finding the most general symmetric form of (B.6) as given in (B.7).  The result 
should be an equation for $E(\sigma)$ that defines a large class of scalar-tensor 
solutions to (2.26) and (2.27) that comprise the theory.\\
\indent  
\textit{Class A Constraints}.\\
\indent
From (B.27) we have
\begin{equation}						
[2\kappa E(\sigma) - 8\pi E^\prime(\sigma)]U^*(\sigma) = [8\pi E(\sigma) - \kappa\kappa_1^{-1}]U^{*\prime}(\sigma)
\end{equation}
\noindent
and employing (B.3) this becomes
\begin{equation}						
[2\kappa E(\sigma) - \sigma E^\prime(\sigma)]U^*(\sigma) = [\sigma E(\sigma) - \kappa_1^{-1}]U^{*\prime}(\sigma)  \quad . 
\end{equation}
\indent
Examination of (B.40) shows that it has the solution 
\begin{equation}						
E(\sigma)U^*(\sigma) = F(\sigma)\sigma^2
\end{equation}
\noindent
provided the following derivative exists
\begin{equation}						
F^\prime(\sigma)=\kappa_1^{-1}\frac{U^*(\sigma)}{\sigma^3} \quad ,  \qquad	\sigma \neq 0	\quad , 
\end{equation}
\noindent
where $U^*(\sigma)$ is defined in (2.19).  Dividing $U^*(\sigma)$ by $\sigma^3$ 
we have
\begin{equation}						
F^\prime(\sigma) = \kappa_1^{-1} [B\sigma^{-3} + \frac{d}{4}T^*\sigma^{-2} + \frac{a}{2!}\sigma^{-1} + \frac{b}{3!} + \frac{c}{4!}\sigma ]  \quad .
\end{equation}
\noindent
Note that $T^*$ actually is a function $T^*(\sigma)$, which was neglected in (B.43) 
and will be discussed further in \S B.5.  Integration of (B.43) gives
\begin{equation}						
F(\sigma) = \kappa_1^{-1} [- \frac{1}{2}B\sigma^{-2} - \frac{d}{4}T^*\sigma^{-1} + \frac{a}{2!} ln \sigma + \frac{b}{3!}\sigma + \frac{1}{2}\frac{c}{4!}\sigma^2 ]
\end{equation}
\noindent
except for several integration factors.  Necessarily, we must assume $T^*=4$ in 
(B.44) or $d=0$ in order to use this relation at all.$^{12}$\\
\indent  
The combination (B.41) and (2.26) results in 
\begin{equation}						
E(\sigma) = \sigma^2 F(\sigma)U^*(\sigma)^{-1}
\end{equation}
\noindent
for use in (2.26) in conjunction with (B.44), and with $T^*=4$ or $d=0$.  
Substituting (B.41), (B.44), and (B.45), the final term in (2.26) becomes
\begin{equation}						
\frac{1}{\sigma}[E(\sigma)g_{\mu \nu}U^*(\sigma) = \sigma F(\sigma)g_{\mu \nu} \nonumber
\end{equation}
\noindent
\begin{equation}						
= \kappa_1^{-1} [- \frac{1}{2}B\sigma^{-2} - \frac{d}{4}T^*\sigma^{-1} + \frac{a\sigma}{2!} ln ~\sigma + \frac{b}{3!}\sigma^2 + \frac{1}{2}\frac{c}{4!}\sigma^2 ]g_{\mu \nu}  \quad .
\end{equation}
\noindent
Simple power counting of mass-dimensions shows immediately that the negative power 
of $\sigma$ makes (B.46) not renormalizable.  See \S B.5 for more.\\
\indent 
\textit{Class B Constraints.}
Relation (B.7) has a limitation, namely that it is a classical tensor.  Such a procedure 
must not destroy the renormalizability of the result in (B.6).  Hence, 
there is an additional, quantum criterion that constrains (B.7).  That is, 
$U^*(\sigma)$ is a quartic potential which is essential to the quantum symmetry 
breaking process in this theory (\S1.2.1), and is renormalizable.  Whatever $E(\sigma)$ is, 
it must not alter the quartic properties that generate the two vacua in Figure 1.\\
\indent  
From the term $\sigma E^\prime(\sigma)U^*(\sigma)$ in (B.40) it is obvious that the 
solution for $E(\sigma)$ now involves a quintic potential $\sigma U^*(\sigma)$.\\
\indent  
First, a quintic potential violates the standard structure of a Lagrangian having 
mass-dimension four.  It now has five and is nonrenormalizable.  Second, there is 
the famous "insolvability of the quintic" theorem due to Galois and Abel.  
(Conceivably, the Galois-Abel theorem has something to do with renormalization 
theory.)\\
\indent
Any polynomial function $E(\sigma)$ in (B.35) that has a positive power of $\sigma$ 
greater that degree $n=1$ will create a quintic potential term which not only is 
nonrenormalizable but is not even solvable according to Galois' theory of groups.  
If $E(\sigma)=\sigma$ then it cancels the $\kappa=\sigma^{-1}$ term in (B.3), and we 
lose relation $\lambda_{Bag}=\hat{\kappa}B$ in (2.5).\\
\indent
It becomes increasingly apparent that if $E(\sigma)\neq 1$, many inconsistencies may 
arise in the theory.  Hence we adopt for (2.26) an arbitrary constant for $E(\sigma)$.  
Since the arbitrary constant can be absorbed into $U^*(\sigma)$, we choose
\begin{equation}						
E(\sigma) = 1 
\end{equation}
\noindent
as the basis for Class B constraints in the present theory.  This appears to satisfy 
both the classical and quantum considerations.  This is also consistent with having 
kept the $\lambda(\phi)$ term on the left-hand side of (2.4), then moving it after 
deriving (2.26) to become a part of $U^*(\sigma)$ -- something that is seen in the 
literature.

\subsection{General constraint for $A(\sigma)$}

\noindent
Note that (B.3) resulted in relation (B.32) in Appendix B.2.  As discussed in 
Appendix B.1, however, $\kappa(\sigma)$ can be any function that results in 
consistent physics.  From (B.25) using (B.33) and (B.34) 
\begin{equation}						
\frac{1}{2} \kappa [A(\sigma) + 4B(\sigma)] = A^\prime(\sigma) + B^\prime(\sigma) 
\end{equation}
\begin{equation}						
\frac{1}{2} \kappa [A(\sigma) - 2A(\sigma)] = A^\prime(\sigma) - \frac{1}{2}A^\prime(\sigma)
\end{equation}
\begin{equation}						
A^\prime(\sigma) = - \kappa A(\sigma)  
\end{equation}
\begin{equation}						
\frac{dA}{A} = - \kappa d\sigma  \quad . 
\end{equation}
\indent
Hence for a general $\kappa(\sigma)$ ansatz we obtain a functional integral for 
$A(\sigma)$
\begin{equation}						
A(\sigma) = e^{-\int \kappa(\sigma)d\sigma}  \quad . 
\end{equation}

\subsection{Complications introduced by the $d\sigma$ term in $U^*(\sigma)$}

\noindent
It needs to be said that the only difference between $U^*(\sigma)$ in (2.19) and 
$U(\sigma)$ in (1.11) is the linear $d\sigma$ term.  Ostensibly there is no reason to 
disregard this term since it is renormalizable and is used in the literature [71].  
It can thus be retained for pion physics in that form, $d\sigma$ ($d\neq 0$).\\
\indent    
What the $d\sigma$ term adds to the theory is to introduce the ability to skew 
(not "tilt") the symmetry breaking potential $U^*(\sigma)$ along the line 
$U^*=–d\sigma$.  In practice, the terms in (2.19) appearing as $B$ and the polynomial 
coefficients $a,b,c,d$ must be adjusted to conform with experimental data.\\
\indent
When coupled to the trace of the energy-momentum tensors, however, it does present 
complications that are discussed below.\\
\indent
\textit{Nonlinear wave equation.}  The $dT^*\sigma$ and $dT^M\sigma$ terms 
necessarily modify field equation (2.27) for $\sigma$.  As mentioned in Appendix B.3, 
it represents at least two additional difficulties.  The equations for $T^*$ and 
$U^*$ are coupled:
\begin{equation}						
U^* = \frac{d}{4}T^*\sigma + U  \quad . 
\end{equation}
\begin{equation}						
T^* = T^M - \sigma_{;\alpha}^2 + 4U^*  \quad . 
\end{equation}
\noindent
which can be uncoupled to give 
\begin{equation}						
U^* = [ \frac{d}{4}(T^M –-  \sigma_{;\alpha}^2)\sigma  + U](1-d\sigma)^{-1}  \quad . 
\end{equation}
\begin{equation}						
T^* = [T^M –-  \sigma_{;\alpha}^2  + 4U](1-d\sigma)^{-1}  \quad . 
\end{equation}
\noindent
and whose derivatives (with respect to $\sigma$) are
\begin{equation}						
U^{*\prime} = [dU^* + \frac{d}{4}(T^M – - \sigma_{;\alpha}^2) + U^\prime](1-d\sigma)^{-1}  \quad , 
\end{equation}
\begin{equation}						
T^{*\prime} = [dT^* + 4U^\prime](1-d\sigma)^{-1}  \quad . 
\end{equation}
\indent
Equation (B.57) for $U^{*\prime}$ must then be substituted into (2.27), yielding a 
highly nonlinear equation of motion due to the $\sigma_{;\alpha}^2$ term which is 
beyond a normal d$\text{'}$Lambertian.  These nonlinearities can be removed by simply setting 
$d=0$.\\
\indent
\textit{Loss of renormalizability.}  The coupling of the $d\sigma$ term ($d\neq 0$) 
to either trace $T^*$ or $T^M$ creates an interaction term $dT^*\sigma$ or 
$dT^M\sigma$ in Lagrangian $\pounds_{G,\sigma}^{(1)}$ (2.16).  From (B.53) and (B.54) this 
produces fifth-degree polynomials $U\sigma$ and $U^*\sigma$ which are not renormalizable by 
simple power counting of mass-dimension.\\
\indent  
Furthermore, a fifth-degree $U\sigma$ or $U^*\sigma$ is a quintic polynomial and is 
subject to the same ``insolvability of the quintic'' theorem due to Galois and Abel 
mentioned in \S B.3 above.  Again the Galois-Abel theorem appears to coincide with the 
loss of renormalizability.

\subsection{Conformal invariance and Jordon-Einstein frames}

\noindent
Conformal transformations
\begin{equation}						
g_{\mu \nu} \rightarrow \tilde{g}_{\mu \nu} = \omega^2(x)g_{\mu \nu}
\end{equation}
\noindent
are relevant to any metric theory of gravity involving $g_{\mu \nu}$, where 
$\omega(x)$ is a non-zero suitably differentiable function of spacetime.  An 
example is Penrose$\text{'}$s conformal mapping technique for visualizing asymptotic 
infinities [59].  (B.59) provides different conformal representations of a given 
Lagrangian such as (B.1) or (B.4), forming a conformal gauge group.  It also has been 
used to generate JFBD solutions from Einstein ones [72].\\
\indent
However, Einstein gravity is not invariant under (B.59).  There are infinitely many 
such conformal frames.  Pauli advised Jordon [67] to be careful about what conformal 
frame was being used.  For example, by setting $\omega^2=f_1(\phi)$ for $f_1(\phi)$ 
in (B.2) or $\omega^2=\phi R$ for $\phi R$ in (B.4), the original JFBD nonminimal 
coupling term can be converted back into the original E-H term with minimal coupling.  
This subject has been reviewed in [62].\\
\indent
Hence, (B.59) creates serious problems of interpretation by producing ambiguities in
 the definition of observables in physics [73].  Selecting the wrong conformal 
representation (or frame) can lead to violations of conservation laws, the equivalence 
principle, and interpretation of experimental results.\\
\indent  
At the quantum level, however, the conformal anomalies break the conformal 
invariance (B.59) of the classical theory.  In QCD$\text{'}$s chiral limit, these set the 
scale for color confinement and hence determine the masses of hadrons (which is 
most of ordinary matter).   Quantum breaking of classical conformal invariance seems 
to resolve this interpretation problem entirely.\\
\indent
In the present report, the original JFBD Lagrangian was described as of the 
Jordan-type and is adopted throughout this study.  That Lagrangian and its equations 
of motion are the Jordan frame.  When $g_{\mu \nu}$ is conformally transformed back into a 
an E-H form, the result is referred to as an Einstein frame in the literature.  
Conformal transformations are not used in this study.

\section{Lagrangians, renormalization, \& lack of consistency}

\subsection{Lagrangians and renormalization}

\noindent
Stelle [74,75] pursued the question of renormalization of the action for quantum 
gravity when it includes terms quadratic in the curvature tensor $R$.  From the 
Riemann tensor $R_{\mu \nu \alpha \beta}$, Ricci tensor $R_{\mu \nu} = 
R_{\mu\alpha\nu\beta} g^{\alpha \beta}$, $R$, and $g_{\mu \nu}$, the following 
action
\begin{equation}						
S = \!\int d^4 x \sqrt{-g} \big[-\frac{1}{2}\kappa^{-1}(R-2\lambda) + \alpha R_{\mu \nu} R^{\mu \nu}  + \beta R^2 \big]
\end{equation}
\noindent
was found to be renormalizable to all orders, with standard assumptions about 
topological and surface terms. For example, some contributions involving the Weyl 
conformal tensor and Gauss-Bonnet invariance vanish.  These actions involve 
fourth-order derivatives and are sometimes referred to as higher-derivative gravity 
or $R^2$-gravity.  The complexity of quantizing fourth-order gravity theory is 
spelled out by Barth \& Christensen [76].\\
\indent
Along this same line, the subject of QFT in curved spacetime has been studied 
extensively by Buchbinder \textit{et al.} [79,23].  Thinking of the covariant 
derivatives in (C.1) as momenta $k$, the consensus of opinion is that quadratic 
$R$-type terms prevail at high energy and strong gravity while the reverse is true 
for the E-H portion of the action at low energy and weak gravity.  Hence (C.1) is an 
effective action in effective QFT with General Relativity as the low-energy Solar 
System limit.\\
\indent  
The shortcomings of (C.1) have also been identified by Stelle.  First is that 
unitarity has been sacrificed.  Higher-derivative Lagrangians manifestly involve the 
propagation of massive Spin-2 ghost states as can be seen by separating into partial 
fractions a typical propagator term: 
$m^2k^{-2}(k^2+m^2)^{-1} = k^{-2} –- (k^2+m^2)^{-1}$.  The minus sign of the second 
term means either a negative energy or a negative norm in state vector space [74].  
Second, the two right-hand higher-derivative terms improve the divergence structure 
of the quantum theory by making (C.1) renormalizable.  Unfortunately, they also 
introduce additonal problems that seem to make the resulting model unsuitable for a 
fundamental theory [75].  These include a new massive graviton plus massive scalar 
excitations which increase the helicity degrees of freedom to eight instead of five 
or at best two.  Negative energy or indefinite norm and third-order time derivatives 
in the Cauchy problem also result.  Stelle$\text{'}$s assessment is that (C.1) seems 
unlikely to find its place in a fundamental theory.\\
\indent  
There is yet another problem with (C.1).  The presence of a dimensional coupling 
constant $\kappa$ in (2.1) is known to be related to the nonrenormalizability of 
Einstein gravity using perturbation theory (an inevitable outcome because Newtonian 
gravity introduces $\kappa$ and must be one limit of GR).   In natural units 
$\hbar=c=1$, the only dimensional quantity is mass $m = (length)^{-1}$ in any given 
action.  Hence a Lagrangian density $\pounds$ appearing in the actions (2.1) and (C.1) 
has a mass-dimension of four because the action $S$ is dimensionless.  In (C.1) $R$, 
$R_{\mu \nu}$, $\kappa^{-1}$, and $\lambda$ have mass-dimension two, while both 
$\alpha$ and $\beta$ are dimensionless.  The standard for a renormalizable 
Lagrangian using perturbation theory, adopted by most authors, is to find an action 
that contains only dimensionless coupling constants such as $\alpha$ and $\beta$, 
while introducing combinations of field terms that have dimensionality four.
The not-so-subtle difference can be seen in the following change to (C.1) [78, 79]
\begin{equation}						
S = \!\int d^4 x \sqrt{-g} \big[-\frac{1}{2}\kappa^{-1}R + \lambda) + \alpha R_{\mu \nu} R^{\mu \nu}  + \beta R^2 \big] 
\end{equation}
\noindent
where the dimensionality of $\lambda$ has changed from two in (C.1) to four in (C.2).  
This simple change addresses a fundamental aspect of the CCP discussed in \S 2.1 
regarding the unification of gravity with QFT.  To add to the confusion, the 
$\lambda$ in Stelle$\text{'}$s original paper was dimensionless because he coupled it using 
$\kappa^{-2}$ [73, p. 962].\\
\indent
Finally, renormalization of JFBD theory has been discussed [81].  It is widely known 
that using a conformal transformation, the JFBD Lagrangian can be changed back into 
Einstein form.  This has been done [81] to``prove'' that JFBD gravity suffers from 
the same nonrenormalizability problems as does (2.1).\\
\indent 
The problem with this argument is that it neglects the fact that conformal 
invariance is broken (lost) at the quantum level due to the conformal anomalies.  
Furthermore, great confusion can arise over the choice of physical frame under 
conformal invariance [73], and this aspect of JFBD Lagrangians is further discussed 
in Appendix B.6.

\subsection{Lack of consistency in QFT}

\noindent
Arbitrary spin in QFT has a long history of pathological problems in the presence of 
interactions.  Higher spin fields ($S > 1$) are also well known to suffer consistency 
problems on curved backgrounds [82,83].\\
\indent  
It was shown years ago that when all allowable interactions under irreducible 
representations of the inhomogeneous Lorentz group are taken into account, QFT is 
not globally well-behaved and exhibits acausal propagation.  The reason is 
straightforward.  The relativistic wave equations always look the same, much like 
the Klein-Gordon equation for a scalar particle $\phi$ or the Dirac equation for a 
fermion $\psi$, with a Lorentz scalar interaction term added to the mass $m$.  For varying 
spins, however, different auxiliary conditions must be built into the wave function 
in order to maintain manifest covariance and Lorentz invariance.  This problem 
actually begins with Spin-0 and applies to all spins, although it does not seem to 
affect Spin-1/2 because the Dirac equation requires no auxiliary components.\footnote{Note 
that a Spin-0 scalar $\phi$ has auxiliary components $\partial_\mu \phi$ in 
Petiau-Duffin-Kemmer theory which are unavoidable.}\\
\indent
The seminal paper was that of Fierz and Pauli [56] who first formulated the problem 
of finding a consistent interactive Lagrangian for Spin-2 massive particles in 
quantum gravity.  Velo and Zwanziger later found the existence of acausal 
propagation for Spin-1 and inconsistencies in Spin-3/2 [84].  Iwasaki [85] 
identified inconsistencies in Spin-2 propagators, noting that the redundant 
components for manifestly covariant wave functions of higher spin are suppressed by 
using subsidiary conditions which the currents coupled to these fields do not 
satisfy.  It has also been claimed that the root of the Spin-2 problem seems to 
have to do with gauge invariance [86].\\
\indent 
Because free and interacting fields generally transform differently under Lorentz boosts, 
it is by no means obvious how to introduce QFT interactions in a Lorentz-invariant manner 
[87].  Effective field theory cannot solve this problem except by breaking Lorentz 
invariance, because if it occurs in one Lorentz frame it occurs in all of them at all 
energies since it involves singularities in Lorentz scalars.  In certain 
circumstances, the Hamiltonian is nonlocal or non-existent in which case one is left only 
with the wave equation for solving energy eigenvalue problems since the Schr\"odinger 
equation is incalculable or does not exist [87].\\
\indent
Although the Weinberg-Witten theorem [88] has cleared up a number of issues regarding 
composite and elementary particles, it has not resolved the inconsistencies in QFT 
discussed in this section.

\subsection{Does a consistent quantum theory of gravity exist?}

\noindent
Since attempts to construct a well-posed unitary, renormalizable QFT for gravity with 
satisfactory Cauchy data to solve problems in physics appear to have failed, the question 
has been raised as to whether a consistent quantum theory of gravity actually exists, and 
if so, what form does it take [89].  The inconsistencies remain a problem and may have 
little to do with gravitation as opposed to the limitations of quantum physics when using 
perturbative methods in nonperturbative regimes [90,75,22].

\section{Classical bag, gravity solutions}

\subsection{Special case classical solutions}

\noindent
Because $\lambda \rightarrow \lambda_{Bag}$ has significantly changed by 44 orders of 
magnitude on going from the bag exterior to the interior, the classical solutions need to 
be addressed.  Bag geometry is assumed to be a sphere.  The bag surface divides space into 
two parts, reminiscent of the Einstein-Straus problem [99] (that is actually concerned with 
the influence of cosmological expansion on an embedded local Schwarzschild metric 
representing the Solar System) [116].  One embeds a Schwarzschild solution into a 
pressure-free expanding Universe by smoothly matching the metrics.\\
\indent  
We find the solution for metric (3.1), considering a static bag in stable equilibrium 
of radius $r=r_{Bag} \neq r_s$ filled with a perfect fluid.  $U(\sigma)$ is 
simplified to $U(\sigma)=B$ and the wave equation for $\sigma$ (2.27)-(2.28) is neglected.  
The hadron mass $m_h$ is assumed to be $m_h= m_h(r)$ and quark charge will be 
discussed.  An objective here is to show the procedure for matching the interior and 
exterior solutions.\\
\indent
Each case will begin with the Einstein-limit solution where $\Omega \rightarrow 
\infty$, following Moller [50].  Approximate solutions from JFBD theory are then 
given for comparison.  Cosmological solutions are relevant but will be addressed in 
another study.\\
\indent
\textit{Case (a).  Exterior vacuum solution ($T^*_{~\mu \nu}=0$).}\\
\indent 
\textit{Case (a.1) – Einstein Gravity ($\lambda=\Lambda=0$).}  Here is the standard 
Schwarzschild exterior problem with $\lambda=0$, for reference.  This means the 
metric is given by (A.2) in Appendix A, with $\lambda=0$ and $\Omega \rightarrow 
\infty$.\\
\indent 
The Schwarzschild solution is well-known, representing the gravitational field of an
object (the hadron) having mass $m=m_h$ extending from the bag surface 
$r=r_s$ to asymptotic infinity with metric solution (A.2) ($\lambda=0$).\\
\indent  
If charged, the Reissner-Nordstr\"om point-mass solution [117] is equally 
well-established for a charge $e$ with electrostatic Coulomb coefficient $k$.  
Putting the two together, (A.3) becomes [118] 
\begin{equation}						
e^\nu  =  1 –-  \frac{2GM}{r}  +  \frac{ke}{r}  –-  \frac{\lambda}{3} r^2 =  e^{- \zeta}   \quad , 
\end{equation}
\noindent
for the exterior gravitational metric in (3.1).  This is the KS metric with a Coulomb 
term $ke^2/r$ added for a charge $e$.\footnote{The charge $e$ can have two 
signs is why the sign changes for the Coulomb term.}\\
\indent
One of the most thorough and straight-forward derivations of the static exterior 
solution uses the parameterized post-Newtonian approximation (PPN) [109] and is 
given by Weinberg [69, p. 244-248].\\
\indent  
Probably the best way to represent this solution is to use Eddington-Robertson 
parameters ($\alpha,\beta,\gamma$ defined below) which collect the answer to second 
and fourth order in the expansion.\\
\indent  
The Reissner-Nordstr\"om analog for a mass $M$ of charge $e$ in the exterior 
($\lambda=0$) is given in [113]
\begin{equation}						
\overbrace{ g_{oo}}^{4} = \frac{2GM}{r} - \frac{4\pi e^2 G}{r^2}\frac{3+2\Omega}{4+2\Omega}
\end{equation}
\begin{equation}						
\overbrace{ g_{oo}}^{4} = - (\gamma - 1 +2\beta) \frac{2G^2M^2}{r^2} - \frac{4\pi e^2 G}{r^2}\frac{3+2\Omega}{4+2\Omega}  \quad . 
\end{equation}
\indent
The exterior point-mass, charged solution is (D1).\\
\indent 
\textit{Case (a.2) – Scalar-Tensor Gravity ($\lambda=\Lambda_{F-L} \neq 0$).}  We 
begin by noting that the vacuum JFBD equations can be written [114]
\begin{equation}						
R_{\mu \nu} = - ~\xi~\sigma_{;\mu} \sigma_{;\nu}  \quad , 
\end{equation}
\noindent
where $\xi$ is proportional to $\kappa_1$.\\
\indent  
Following the same PPN procedure, the result is ($\alpha \equiv 1, \beta=1$)
\begin{equation}						
\overbrace{ g_{oo}}^{2} = \frac{2GM}{r}
\end{equation}
\begin{equation}						
\overbrace{ g_{oo}}^{4} = - (\gamma - 1 +2\beta) \frac{2G^2M^2}{r^2} 
\end{equation}
\begin{equation}						
\overbrace{ g_{ij}}^{2} =  (3\gamma - 1)\delta_{ij} \frac{GM}{r} + (1 - \gamma)\frac{GMx_i x_j}{r^3} 
\end{equation}
\noindent
where
\begin{equation}						
\gamma = \frac{\Omega + 1}{\Omega + 2}  \quad . 
\end{equation}
\noindent
$\Omega$ appears at fourth order (D.6) and in the off-diagonal mixing terms at 
second order (D.7).  There is also a third-order spin-orbit coupling effect on the 
precession of perihelia which is not shown.  The $\gamma$-term in (D.6) was found 
earlier in [112].\\
\indent
For a static spherically symmetric mass, both JFBD and Einstein gravity exhibit the 
property that the gravitational field depends on $M$ but not any other property of 
the mass.\\
\indent
\textit{Case (b). Interior solution ($\lambda=\lambda_{Bag}= \hat{\kappa}B$ and 
$T^*_{~\mu \nu} \neq 0$).}\\
\indent
We now want to give the mass a finite spatial extent, forming a static bag of radius 
$r=r_s$ with $r \leq r_s$.\\
\indent
\textit{Case (b.1) – Einstein Gravity.}  We work in the Einstein limit 
$\Omega \rightarrow \infty$ to demonstrate the method.  Consider a general case such 
as $N_q=2$ quarks. Note that this is an intractable nonlocal 3-body (or more) 
problem.\\
\indent
To find the interior solution, match the metrics at the surface, assume the pressure 
is zero there, and solve for the final answer.\\
\indent 
The interior solution for (3.1) is not (D.1) but [50]
\begin{equation}						
e^\nu = (A - Ce^\zeta) 
\end{equation}

\begin{equation}						
e^{-\zeta} = (1 - r^2/R^2)^{- 1/2} 
\end{equation}

\begin{equation}						
R^2 = 3/(\lambda + \kappa \rho) 
\end{equation}

\begin{equation}						
A = (1 - r_s^2/R^2)^{-1/2} 
\end{equation}

\begin{equation}						
C = 1/2  \quad . 
\end{equation}

Using (3.2), (D.11) becomes
\begin{equation}						
R^2 = 3/\kappa(\rho - B)  \quad . 
\end{equation}
\noindent
The pressure equation is given by [50, Moller]
\begin{equation}						
\kappa p = \frac{3Ce^{2\zeta} - A}{R^2e^\nu} + \lambda  \quad . 
\end{equation}
\indent
Consider first the case of zero charge ($e=0$).  One adjusts the constants $A$ and 
$C$ so that (D.9)-(D.14) and (D.1)-(D.3) coincide at the surface $r= r_s$.  Also $p$ in 
(D.15) has to be zero, $p=0$.  These conditions then lead to the following solutions for 
$A$ and $C$ in (D.9),
\begin{equation}						
A = \frac{3}{2} (1 - r_s^2/R^2)^{+1/2} 
\end{equation}
\begin{equation}						
C = 1/2  \quad . 
\end{equation}
\noindent
along with the mass relation
\begin{equation}						
m_h(r) = \frac{1}{2}r_s^3/R^2 = \frac{1}{6}(\lambda + \kappa \rho)r^3  \quad ,~~r<r_s~~. 
\end{equation}
Inserting (3.2) gives
\begin{equation}						
m_h(r) = \frac{1}{6}\hat{\kappa}(\rho - B)r^3  \quad ,~~r<r_s~~, 
\end{equation}	
\noindent
inside the hadron bag, and 
\begin{equation}						
m_h = \frac{4}{3}\pi r_s^3 \rho  \quad ,~~r>r_s 
\end{equation}
\noindent
outside the hadron when $\lambda \equiv \Lambda =0$ in the exterior.\footnote{Mass $M$ in 
Einstein gravity is $M=4\pi \int r^2e^{-\zeta}dV$ and is metric dependent.  It is an 
asymptotic concept and only equals $\rho V$ for a volume $V=4\pi r^3/3$ in special 
circumstances.  See Moller [50].  For the Arnowitt-Deser-Misner mass in the Jordan frame, 
it is $M=4\pi\int \rho r^2dr$.}\\
\indent  
For a nonzero charge ($e \neq 0$), a Coulomb term can be added just as it was 
introduced into (D.1), and this procedure followed.\\
\indent
\textit{Case (b.2) – Scalar-Tensor Gravity.}  As mentioned earlier, this is the more 
complicated solution that is new when $\lambda\!\neq\!0$ in the interior.\\
\indent  
The case of an interior solution for JFBD has already been developed in [113].  The 
problem with that work is that it assumed $\lambda=0$.  The solutions along with the 
pressure $p$ and scalar $\sigma$ are
\begin{equation}						
g_{oo} =  - 1 + \frac{r_o^2}{R^2}\Big(\frac{3\Omega + 7}{3 + 2\Omega}\Big) - \Big(\frac{r^2}{R^2} \frac{\Omega + 3}{3 + 2\Omega}\Big)
\end{equation}
\begin{equation}						
g_{rr} =  1 - \frac{r_o^2}{4R^2}\Big(\frac{6\Omega + 19}{3 + 2\Omega}\Big) + \Big(\frac{r^2}{4R^2} \frac{6\Omega + 15}{3 + 2\Omega}\Big) 
\end{equation}
\begin{equation}						
p = \rho \Big[ \frac{r_o^2}{2R^2}\Big(\frac{\Omega + 3}{3 + 2\Omega}\Big) - \Big(\frac{r^2}{2R^2} \frac{\Omega + 3}{3 + 2\Omega}\Big) \Big] 
\end{equation}
\begin{equation}						
\sigma = \sigma_o \Big[ 1 + \Big(\frac{r_o^2}{2R^2(3 + 2\Omega}\Big) + \Big(\frac{r^2}{2R^2(3 + 2\Omega}\Big) \Big] 
\end{equation}
where $R^2$ is given by (D.14).  Again, $\lambda=0$ and $a\!=b\!=c\!=f\!=0\!$ was 
assumed in (2.22) and (2.24) in order that the authors of [113] could arrive at (D.24).\\
\indent
A thorough discussion of the interior solutions and other features of the 
scalar-tensor solutions will be presented elsewhere.\\
\indent
\textit{Case (b.3) - Introducing $\lambda$.}  Recalling that all of Appendix D 
addresses approximations without $\lambda$, we are now prepared to include $\lambda$ using 
a very simple trick.\\
\indent  
By virtue of the bag constant $B$ having been introduced as a negative pressure in 
(3.2) and (D.14), solution (D.21)-(D.24) where $\lambda$ was assumed to be zero can be 
converted into a next-order approximation that now includes the vacuum energy density 
within the bag.  That is, $B$ is in the solution (D.21)-(D.24) via (D.14).  For that matter, 
so is $G \sim \sigma^{-1}$.  The one exception is the $\sigma$ solution in (D.24) due 
to [113] which we know is incorrect since that derivation assumed that $m_\sigma$ 
was zero.  Here, on the other hand, a scalar mass $m_\sigma = \sqrt{a}$ is 
introduced from (B.38).  Hence, the solution in (D.24) has to be cut off by an 
exponential damping factor $\sigma \sim \sigma_o (e^{-\mu r})$ at the bag$\text{'}$s 
surface $r=r_s$ as mentioned in the discussion of (2.28).\\
\indent
What also is new is the equation of motion (2.27) for the $\sigma$-field, which is 
highly nonlinear and is driven by two sources, the trace of the matter 
energy-momentum tensor $T^M$ (e.g. [70] for driving variations in $G$) and the 
$f$-coupling to the quarks $\psi$.\\
\indent
\textit{Case (c).  Quark-gluon dynamics.}  Most importantly, there is nothing in the 
dynamics of the scalar-tensor model presented here that limits the basic results of 
NTS bag theory (1.13) which intrinsically includes QCD in the exact gluon limit, 
$\pounds_{NTS} \rightarrow \pounds_{QCD}$ (except that the quarks follow geodesics).  
A mean-field approximation (MFA) [119,19] for the charged interior is a feasible 
means for addressing quark dynamics within the model here.\\
\indent
\textit{Summary.}  There are two mass mechanisms involved, a graviton mass $m_g$ 
associated with $\lambda$ in (2.31)-(2.34) and Appendix A, and an NTS mass associated with 
the $\sigma$-field $m_\sigma$ in (2.22) and (2.24)) and Appendix C.2 derived from JFBD theory.  
Both introduce short-range behavior when $U^*(\sigma)$ or $U(\sigma)$ breaks the 
symmetry of the vacuum.

\subsection{Asymptotic freedom \& short-range gravity}

\noindent
In the classical, weak-field Newtonian approximation for spherically symmetric 
metrics such as Kottler-Schwarzschild (KS) in Appendix A, the coefficient $e^\nu$ of 
$dt^2$
\begin{equation}						
e^\nu = 1 - \frac{2M}{r} - \frac{\lambda}{3}r^2 
\end{equation}
\noindent
is equivalent to $(1 - 2\Phi_N)$ where $\Phi_N=GM/r$ is the effective Newtonian 
potential.  This gives a gravitational acceleration $\ddot{r} = – \nabla_r \Phi_N = 
GM/r^2$ when $\lambda=0$.\\
\indent  
Following an early argument by Freund \textit{et al.} [120,121] for $\lambda \neq 0$, 
the effective Newtonian potential is actually
\begin{equation}						
\Phi_N  =  - \frac{GM}{r} - \frac{1}{6}\lambda r^2  \quad . 
\end{equation}
\noindent					
$\lambda \neq 0$ represents a harmonic oscillator potential superposed on the 
Newton law, and is the source of a non-Newtonian force in the Newtonian limit.  In 
de Sitter space ($M=0$), (D.26) becomes
\begin{equation}						
\Phi_N  =  - \frac{1}{6}\lambda r^2  \quad , 
\end{equation}
\noindent
giving 
\begin{equation}						
\ddot{r}  =  -\nabla_r \Phi_N  =  1/3 \lambda r 
\end{equation}
\noindent
which is a harmonic oscillator equation, depending on the sign of $\lambda$.  Based 
upon the tachyonic graviton mass argument against a negative $\lambda$ (\S 2.4), 
(D.28) has the wrong sign.\\
\indent
It is the hadron interior that needs to be addressed.  Recalling the interior 
solution (D.9)-(D.15) with the radial mass dependence $m(r)$ in (D.18), there is a 
different answer.\\
\indent
First (D.9) gives
\begin{equation}						
e^{2\nu}  =  (A - Ce^\zeta)^2  =  A^2 - 2ACe^\zeta - C^2e^{2\zeta}  \quad . 
\end{equation}
\noindent
In the weak-field Newtonian approximation $g_{oo}= e^{2\nu} \approx (1 –- 2\Phi_N)$, 
one determines that 
\begin{equation}						
\Phi_N  =  \frac{1}{2}(A^2 - \frac{3}{4}) - \frac{1}{2}A\sqrt{1 - r^2/R^2} - \frac{1}{8}r^2/R^2  \quad . 
\end{equation}
\noindent
This yields a radial force per unit mass $\ddot{r} = –\nabla_r \Phi_N$ of
\begin{equation}						
\ddot{r}  =  \frac{1}{R^2}[ \frac{1}{4} - \frac{A}{\sqrt{1-r^2/R^2}} ]r 
\end{equation}
\noindent
within the hadron bag.  (D.31) vanishes when $4A=e^\zeta$ or $r=0$.  When $4A>e^\zeta$ 
and $r$ sufficiently small, (D.31) is a simple harmonic oscillator equation that is 
not dependent upon the sign of $\lambda$.  $R^2$ is given by (D.14), and $A$ by (D.12).  
Hence the $r^2/R^2$ term is negative as long as $B>\rho$ preventing the radical from 
being imaginary.  As a cautionary note, (D.31) is a classical approximation that 
requires renormalization features.\\
\indent  
Without any further assumptions, the NTS scalar-tensor model has provided a quark 
potential term that is consistent with those using QCD potential models such as 
[122,34] 
\begin{equation}						
V(r)  =  \alpha r + \beta r  \quad . 
\end{equation}
\noindent
In the present model, (D.31) introduces a natural contribution to $\alpha$ in (D.32) 
determined by the bag constant $B$ which is a negative pressure.\\
\indent
The discovery of asymptotic freedom [123], the QFT property in QCD that quark and 
gluon interactions weaken at shorter distances, allows for the calculation of 
cross-sections using parton techniques.  In addition to bag models, there are also 
potential techniques such as (D.32), but (D.31) and (D.32) now merge these together.\\
\indent
Of course, (D10) and (D.31) are a gravitational contribution to the QCD color force 
attraction.  Yet these calculations seem to be compatible with current ideas about 
asymptotic freedom [123].

\end{document}